\documentclass[pre,reprint,superscriptaddress,floatfix,aps]{revtex4-2}
\usepackage{amsmath}
\usepackage{amsfonts}
\usepackage{graphicx}
\usepackage{physics}
\usepackage[usenames]{color}

 \def\be{\begin{equation}} \def\ee{\end{equation}}
\def\bea{\begin{eqnarray}} \def\eea{\end{eqnarray}}

\begin{document}

\date{\today}
\title{Prethermalization in aperiodically driven classical spin systems}
 \author{Sajag Kumar}
\email{sajag.kumar@niser.ac.in}
\affiliation{School of Physical Sciences, National Institute of Science Education and Research, a CI of Homi Bhabha National Institute, Jatni 752050, India}
\author{Sayan Choudhury}
\email{sayanchoudhury@hri.res.in}
\affiliation{Harish-Chandra Research Institute, a CI of Homi Bhabha National Institute, Chhatnag Road, Jhunsi, Allahabad 211019}
	
\begin{abstract}
Periodically driven classical many-body systems can host a rich zoo of prethermal dynamical phases. In this work, we extend the paradigm of classical prethermalization to aperiodically driven systems. We establish the existence of a long-lived prethermal regime in spin systems subjected to random multipolar drives. We demonstrate that the thermalization time scales as $(1/T)^{2n+2}$, where $n$ is the multipolar order and $T$ is the intrinsic time-scale associated with the drive. In the $n \rightarrow \infty$ limit, the drive becomes quasi-periodic and the thermalization time becomes exponentially long ($\sim \exp(\beta/T)$). We further establish the robustness of prethermalization by demonstrating that these thermalization time scaling laws hold for a wide range of initial state energy densities. Intriguingly, the thermalization process in these classical systems is parametrically slower than their quantum counterparts, thereby highlighting important differences between classical and quantum prethermalization. Finally, we propose a protocol to harness this classical prethermalization to realize time rondeau crystals.

\end{abstract}
	
\maketitle
\section{Introduction} 
The non-equilibrium dynamics of driven many-body systems have been intensely investigated in recent years~\cite{bukov2015universal,oka2019floquet,moessner2017equilibration,rudner2020band,holthaus2015floquet,eckardt2017colloquium,haldar2022statistical,sen2021analytic}. These systems provide a fertile arena for both Hamiltonian engineering and the realization of intrinsically non-equilibrium phases of matter that do not have any equilibrium analog~\cite{sacha2018review,sacha2020book,khemani2019review,nayak2019review,sondhi2016prb,sondhi2016prl,nayak2016prl,yao2017prl,monroe2017nature}. Unfortunately, due to the absence of any conservation laws, driving inevitably leads to unbounded heating, thereby posing a major challenge to these experiments~\cite{d2014long,lazarides2014equilibrium,choudhury2014stability,choudhury2015transverse,bilitewski2015scattering,bilitewski2015population}.

While it is very difficult for driven systems to evade an ultimate heat death, it is possible to delay this thermalization process significantly. For periodically driven (Floquet) systems, this can be achieved by tuning the drive frequency to a value that is much larger than the local energy scales in the system~\cite{abanin2017effective,morningstar2023universality,ho2023quantum,machado2019exponentially,machado2020long,else2017prethermal,kyprianidis2021observation,o2024prethermal}. In this case, after an initial transient period, the system enters a `prethermal' state, where it does not absorb energy for exponentially long times. Interestingly, this phenomenon of Floquet prethermalization persists both in the classical and quantum regimes.

Recently, the notion of prethermalization has been extended beyond the Floquet paradigm. The most well-studied example of this is the case of quasi-periodic driving, where a long-lived prethermal regime has been theoretically predicted~\cite{dumitrescu2018logarithmically,else2020long,nandy2017aperiodically,nandy2018steady,choudhury2021self,chen2024multifractality,gallone2023prethermalization} and experimentally realized~\cite{he2023quasi,he2024exp,dumitrescu2022dynamical} for a large class of quantum many-body systems. While this is a promising direction, just like the Floquet case, even quasi-periodic driving is completely deterministic. Intriguingly, some recent studies have shown that prethermal phases of matter can also emerge in noisily driven quantum systems as long as the noise is temporally correlated. In particular, prethermalization has been demonstrated for a special class of structured random drives dubbed `random multipolar drives' (RMD)~\cite{zhao2021random,mori2021rigorous,zhao2023temporal,yan2024prethermalization}. As the name suggests, a RMD is characterized by $n-$multipolar correlations, where the $n = 0$ and $n \rightarrow \infty$ limits correspond to a completely random and a quasiperiodic Thue-Morse drive respectively. For any finite integer $n \ge 1$, the prethermalization lifetime scales as $(1/T)^{2n+1}$, where $T$ is a natural time-scale of the drive as explained below. Moreover, in the Thue-Morse limit, the thermalization lifetime scales faster than any power law as $\exp [(\log(1/T))^2]$~\cite{mori2021rigorous}. This multipolar driving protocol has been recently employed to realize a prethermal non-equilibrium phase of matter called the `time rondeau crystal' in a $^{13}{\rm C}$-nuclear-spin diamond quantum simulator~\cite{moon2024experimental}. A natural question immediately arises in this context: what is the fate of this prethermalization in the classical regime?

\begin{figure}
    \centering
    \includegraphics[width=0.47\textwidth]{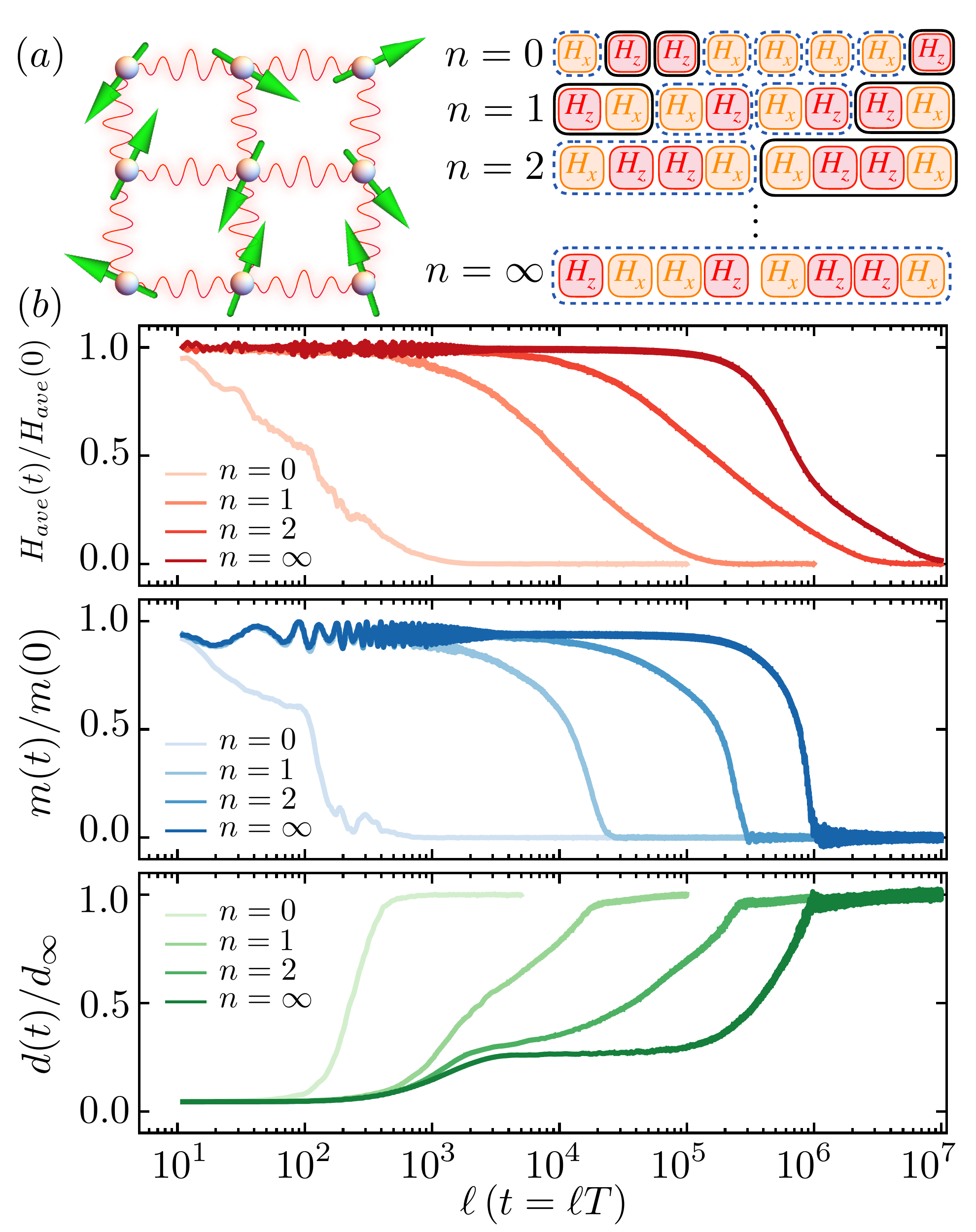}
    \caption{{\bf Model and Dynamics:} (a) Left: Schematic illustration of a system of classical spins on a square lattice with nearest-neighbor Ising interactions. Right: The time evolution of the system is governed by an $n-$random multipolar driving (RMD) sequence of two Hamiltonians, $H_x$ and $H_z$. The explicit form for $H_x$ and $H_z$ is given in Eqs.~\ref{hz} and \ref{hx}. Each $n$-RMD is composed of a random sequence of two blocks of size $2 n$. These blocks are created by concatenating the fundamental blocks of the $(n-1)-$RMD sequence in the two possible ways. The quasi-periodic self-similar Thue-Morse sequence emerges in the $n \rightarrow \infty$ limit. (b) The evolution of the energy density, $H_{\rm ave} = (H_z+H_x)/2$, staggered magnetization, $m = \sum_{ij} (-1)^{(i+j)}S_{ij}$, and the decorrelator, $d(t)$ (see eq.~\ref{decorrelator}) for various $n-$RMDs, clearly establishes the existence of a prethermal regime.}
    \label{fig:illustration}
\end{figure}

This work provides unequivocal numerical evidence for a long-lived prethermal state for RMD systems in the classical regime. Strikingly however, the lifetime of this prethermal regime scales as $(1/T)^{2n+2}$, when the system is initially prepared in a state with ferromagnetic (or anti-ferromagnetic) order. This situation is even more dramatic in the Thue-Morse regime, where the prethermalization lifetime scales exponentially as $(\exp(\beta/T))$. In this context, it is worth noting that classical systems are generically expected to be more chaotic than the corresponding quantum systems due to the absence of any Lieb-Robinson bounds~\cite{mori2018floquet}. This is evident in the growth of the decorrelator (and the corresponding decline in magnetic ordering) at short times (see Fig.~\ref{fig:illustration}). Intriguingly, however, the decorrelator plateaus after the initial growth and the thermalization time is parametrically longer than the corresponding quantum system.  Our results highlight interesting differences between aperiodically driven classical and quantum systems. 

This paper is organized as follows. We introduce the model in sec.~\ref{sec:model}. The dynamics of prethermalization under random multipolar drives is investigated in sec.~\ref{sec:RMD}. We demonstrate that this classical RMD prethermalization can be harnessed to realize time rondeau crystals in sec.~\ref{sec:RMD}. We conclude with a summary of our work and outline some interesting future directions in sec.~\ref{sec:conclusion}.

\section{Model}
\label{sec:model}
We consider a system of nearest-neighbor interacting classical spins $\vec{S}_{ij} = \left(S_{ij}^{x},S_{ij}^{y},S_{ij}^{z}\right) \in {\mathcal S}^2$ on a square lattice of linear size $N = 50$. The time evolution of this system from time, $t =(k-1)T$ to $t= k T$ ($k \in \mathbb{Z}^{+}$) is governed by a Hamiltonian $H_k$, where $H_k \in \{H_z, H_x\}$ is the $k$-th element of a sequence of Hamiltonians. The two distinct elements of the sequence, $H_z$ and $H_x$, are:
\begin{eqnarray}
    \label{hz}
    H_z &=& \sum^{N}_{i,\, j = 1} \left( S^z_{i\,j} S^z_{i+1\,j} + S^z_{i\,j} S^z_{i\,j+1}\right) + hS^z_{ij}, \\
    \label{hx}
    H_x &=& \sum^{N}_{i,\, j = 1} gS^x_{ij},
\end{eqnarray}
where $h$ and $g$ denote the longitudinal and transverse magnetic field strengths respectively. The procedure to generate the sequence that determines $H_k$ is illustrated in fig.~\ref{fig:illustration}(a) and discussed in detail in the next section.

We compute the spin dynamics by integrating the standard equations of motion $\partial_t S_{ij} = \{S_{ij}, H\}$, where $\{\ldots\}$ indicate Poisson brackets, and the spins $S_{ij}$ satisfy the relation $\{S_{ij}^{\alpha},\, S_{i^{\prime}j^{\prime}}^{\beta}\} = \delta_{ii^{\prime}}\delta_{jj^{\prime}}\varepsilon^{\alpha\beta\gamma}S_{ij}^{\gamma}$. Following Howell {\it et al.}~\cite{howell2019asymptotic}, we analytically integrate these equations to obtain the following stroboscopic time evolution:

\begin{equation}
\label{eq:dynamics}
    \Vec{S}_{ij}(\ell T + T) = 
    \begin{cases}
       R_x^{ij}\Vec{S}_{ij}(\ell T), &\text{if}\,\, H(\ell t) = H_x\\ R_z^{ij}\Vec{S}_{ij}(\ell T), &\text{if}\,\, H(\ell t) = H_z
     \end{cases}.
\end{equation} 
Here, $R_x$ and $R_z$ correspond to the following rotation operators about the $x$ and $z$ axis:
\begin{eqnarray}
    R_x^{ij} &=& \begin{pmatrix}
1 & 0 & 0 \\
0 & \cos \left( g T\right) & -\sin \left( g T\right) \\
0 & \sin \left( g T\right) & \cos \left( g T\right) \\
\end{pmatrix}, \\
R_z^{ij} &=& \begin{pmatrix}
\cos \left( \kappa_{ij} T\right) & -\sin \left( \kappa_{ij} T\right) & 0 \\
\sin \left( \kappa_{ij} T\right) & \cos \left( \kappa_{ij} T \right) & 0 \\
0 & 0 & 1 \\
\end{pmatrix},
\end{eqnarray}
where $\kappa_{ij} = (S^{z}_{i+1\,j} + S^{z}_{i-1\,j} + S^{z}_{i\,j+1} + S^{z}_{i\,j-1}) + h$ is the effective magnetic field along the $z$-direction. We note that the rotation $R_z$ is non-linear, thereby leading to chaotic dynamics~\cite{howell2019asymptotic,mori2018floquet}.

It is already known that this system exhibits prethermalization for a periodic driving protocol composed of an alternating sequence of $H_z$ and $H_x$. This prethermalization is a consequence of drive-induced synchronization and it can be leveraged to realize classical prethermal phases of matter like discrete time crystals~\cite{pizzi2021classicalPRL,pizzi2021classical,ye2021floquet}. In the remainder of this work, we systematically analyze the dynamics of this system under different sequences generated by $H_z$ and $H_x$. \\

\begin{figure*}
    \centering
    \includegraphics[width=0.99\textwidth]{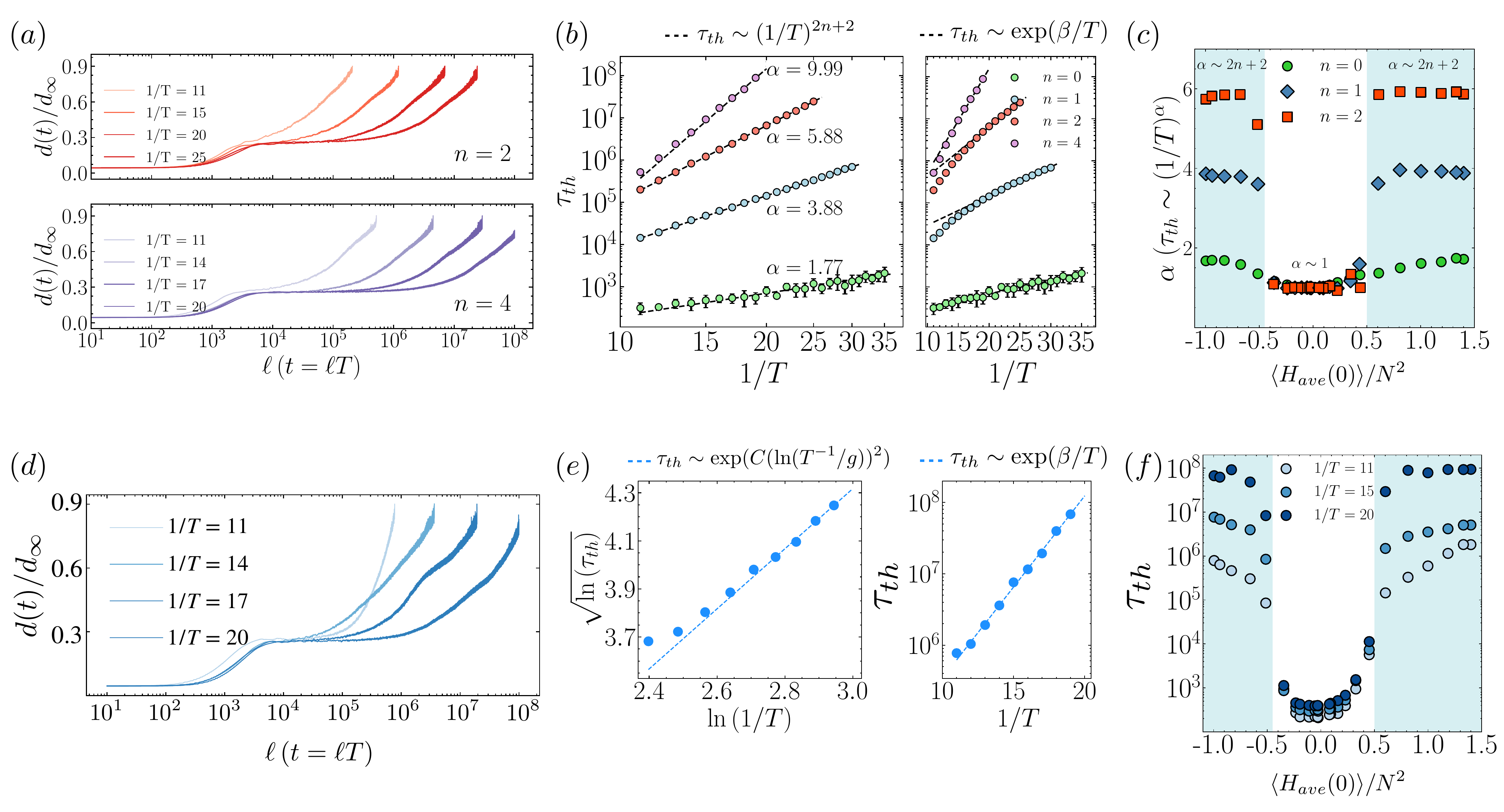}
    \caption{{\bf Top Panel: RMD prethermalization}: (a) The time-evolution of the decorrelator, $d(t)$ (eq.~\ref{decorrelator}) for different driving frequencies for random quadrupolar driving (top), and random sextapolar driving (bottom).  A long-lived prethermal regime is seen in both cases. (b) The thermalization time, $\tau_{\rm th}$ (see text) is fit to a power law (left) and an exponential (right). The power law fit ($\tau_{\rm th} \sim (1/T)^{(2n+2)}$) is better than the exponential fit. The error bars are obtained by averaging over $(20, 10, 5, 1)$ different cycles for $n = (0, 1, 2, 4)$. (c) The robustness of prethermalization is established by examining the dependence of the $\tau_{\rm th}$ scaling exponent, $\alpha$ on the energy density of the initial state. We observe that $\alpha \sim 2n+2$ for both highly positive and negative energy states; for other initial states, the system thermalizes rapidly with the same value of $\alpha$ for all RMDs. {\bf Bottom Panel: Thue-Morse prethrmalization:} (d) A prethermal plateau is clearly observed in the time evolution of the decorrelator for different driving frequencies (e): The fit of the thermalization times to the scaling function, $\tau_{\rm th}$ to the function $\exp \left(C(\ln(T^{-1}/g))^2\right)$ (left) and $A \exp(\beta/T)$. The numerical evidence clearly indicates that $\tau_{\rm th}$ scales exponentially with the driving frequency, much like the case of perfectly periodic driving. (f) The dependence of $\tau_{\rm th}$ on the initial state energy density, $\langle H(0)\rangle /N^2 $. Much like the RMD case, $\tau_{\rm th}$ is long for both highly positive and negative energy states.
    }
    \label{fig:reps_and_scaling}
\end{figure*}

\section{Random Multipolar Drives} 
\label{sec:RMD}
We now proceed to go beyond the periodic driving regime by exploring the time evolution of this system under $n$-RMDs. For $n=0$, the drive sequence is generated by randomly selecting $H_z$ or $H_x$, and it is thus completely devoid of any structure. For $n \ge 1$, the \emph{n-th} RMD is generated by a random array of two $n$-polar blocks, where each such block is obtained by concatenating the two $(n-1)$-polar blocks. To unpack this definition, we first examine the case for $n=1$, where the drive is characterized by dipolar correlations. In this case, the drive is generated by a random sequence of one of two possible blocks: $(H_z,\, H_x)$ or $(H_x,\, H_z)$. Similarly, for $n=2$, the drive is generated by a random array of two quadrupolar blocks: $(H_z,\, H_x,\,H_x,\, H_z)$ or $(H_x,\, H_z,\, H_z,\, H_x)$. In the $n\to\infty$ limit, this procedure yields the self-similar quasiperiodic Thue-Morse sequence.

We begin by examining the time-evolution of the system when it is initially prepared in an N\'{e}el ordered state with spins polarized along the positive z-axis in one sublattice and the negative z-axis in the other. Thus, each spin can be parametrized by two angles $\theta$ and $\phi$ in the form $\Vec{S}_{ij} = \left(\sin(\theta_{ij})\cos(\phi_{ij}),\sin(\theta_{ij})\sin(\phi_{ij}),\cos(\theta_{ij})\right)$. Following previous works on periodically driven systems~\cite{howell2019asymptotic,pizzi2021classicalPRL},  $\theta$ is chosen from a Guassian distribution (with mean $0$ and standard deviation $2\pi W$, and $\phi$ is chosen from a uniform distribution between $0$ and $2\pi$ respectively. When $W=0$, all the spins are perfectly aligned along the positive-$z$ direction in one sub-lattice and the negative-$z$ direction along the other. In this case, the evolution is completely determined by the mean-field Hamiltonian. To incorporate the many-body nature of the system, we set $W \ne 0$. We note that although the results reported in the main text are for $W=0.01$, our results also hold in the $W=0$ limit~\cite{suppmat}. To connect to previous results on Floquet prethermalization~\cite{howell2019asymptotic}, we take $(g, h) = (0.9045,\, 0.809)$.

We characterize the thermalization time-scale of this system by examining the growth of a classical out-of-time-ordered correlator, (a {\emph{ decorrelator}})~\cite{das2018light,bilitewski2018temperature,bilitewski2021classical}:

\begin{equation}
    \label{decorrelator}
    d(t) = \sqrt{\frac{1}{N}\left( \sum^N_{i,\, j = 0} \left( \vec{S}_{ij} - \vec{S}^{\prime}_{ij} \right)^2 \right)},
\end{equation}
where, $\Vec{S}^{\prime}_{ij}$ is obtained by adding a slight perturbation to the original spin $\Vec{S}_{ij}$. The decorrelator thus quantitatively captures one of the most crucial characteristics of chaotic dynamics - the sensitive dependence of initial conditions. We note that the decorrelator has been successfully employed to capture thermalization in driven many-body systems~\cite{pizzi2021classicalPRL,pizzi2021classical,jin2024floquet}. In particular, $d(t)$ saturates to a characteristic value of $\sqrt{2}$, when the driven system reaches an infinite temperature state characterized by trivial correlations~\cite{pizzi2021classicalPRL}. Thus, $d(t)$ plays a role similar to the entanglement entropy in driven quantum systems. For our calculations, $\Vec{S}^{\prime}_{ij}$ has been obtained by adding $2 \pi \Delta \delta$ to both the azimuthal and polar angles of $\Vec{S}_{ij}$; here, $\Delta = 0.01$ and $\delta$ is a standard normal random number. As stated earlier, the complete thermalization of the system to an infinite temperature state is signalled by the saturation of the decorrelator to $d_{\infty} = \sqrt{2}$~\cite{pizzi2021classicalPRL}. For our calculations, we have obtained the thermalization time, $\tau_{\rm th}$ by averaging the times at which $d(t)/d_{\infty} = 0.90, \, 0.89, \, {\rm and} \, 0.88$. Our results are shown in fig.~\ref{fig:reps_and_scaling}. It is clear from these calculations that a long-lived prethermal phase indeed appears for $n-$RMDs, with a thermalization time that scales algebraically with $n$ as $\tau_{th}\sim (1/T)^{2n+2}$; this conclusion does not depend on the exact threshold value of $d(t)/d_{\infty}$~\cite{suppmat}. We emphasize that this classical thermalization time is parametrically longer than the corresponding quantum model ($\tau_{th}^{\rm quantum} \sim (1/T)^{2n+1}$), despite the absence of any Lieb-Robinson bounds on the propagation of information.\\

We now proceed to analyze this prethermalization further by determining the dependence of $\tau_{th}$ the initial state energy density, $\langle H_{\rm ave}(0)\rangle/N^2$, where $H_{\rm ave} = (H_z+H_x)/2$. The procedure to tune $\langle H_{\rm ave}(0)\rangle/N^2$ is detailed in the supplementary material~\cite{suppmat}. As shown in fig.~\ref{fig:reps_and_scaling}(c), we find that the system exhibits a strong dependence of the thermalization time on the initial state energy density; this is a salient feature of prethermalization. Notably, we find that an algebraically long prethermalization exists, when $\langle H_{\rm ave}(0)\rangle/N^2 \le \epsilon_c $ and  $\langle H_{\rm ave}(0)\rangle/N^2 \ge \epsilon^{\prime}_c $, where $\epsilon_c \sim -0.5$ and $\epsilon^{\prime}_c \sim 0.5$. These results demonstrate the robustness of this classical RMD prethermalization. Furthermore, we note that $\tau_{th}$ is small and $n-$independent, when the initial energy density is close to zero. This rapid thermalization occurs due to the proximity of the initial state to the final thermal state.

Finally, we analyze the fate of this prethermalization in $n \rightarrow \infty$ limit, where the driving protocol is described by the completely deterministic Thue-Morse sequence. By examining the decorrelator, we find that the thermalization time grows exponentially with the frequency, $\tau_{\rm th} \sim \exp(\beta/T)$ (see fig.~\ref{fig:reps_and_scaling}(e)); this behavior is strikingly different from the thermalization time in spin-$1/2$ systems, where $\tau_{\rm th} \sim  \exp\left(C (\ln(T^{-1}/g))^2\right)$. Thus, akin to the RMD, Thue-Morse driving also leads to a parametrically longer-lived prethermal regime in classical spin models, compared to their quantum counterparts. We also examine the dependence of $\tau_{\rm th}$ on the initial state energy density and find that a long-lived prethermal regime exists, when $\langle H_{\rm ave}(0)\rangle/N^2 \le \epsilon_c $ and  $\langle H_{\rm ave}(0)\rangle/N^2 \ge \epsilon^{\prime}_c $, where just like the RMD case, $\epsilon_c \sim -0.5$ and $\epsilon^{\prime}_c \sim 0.5$. \\

\begin{figure}
    \centering
    \includegraphics[width=0.45\textwidth]{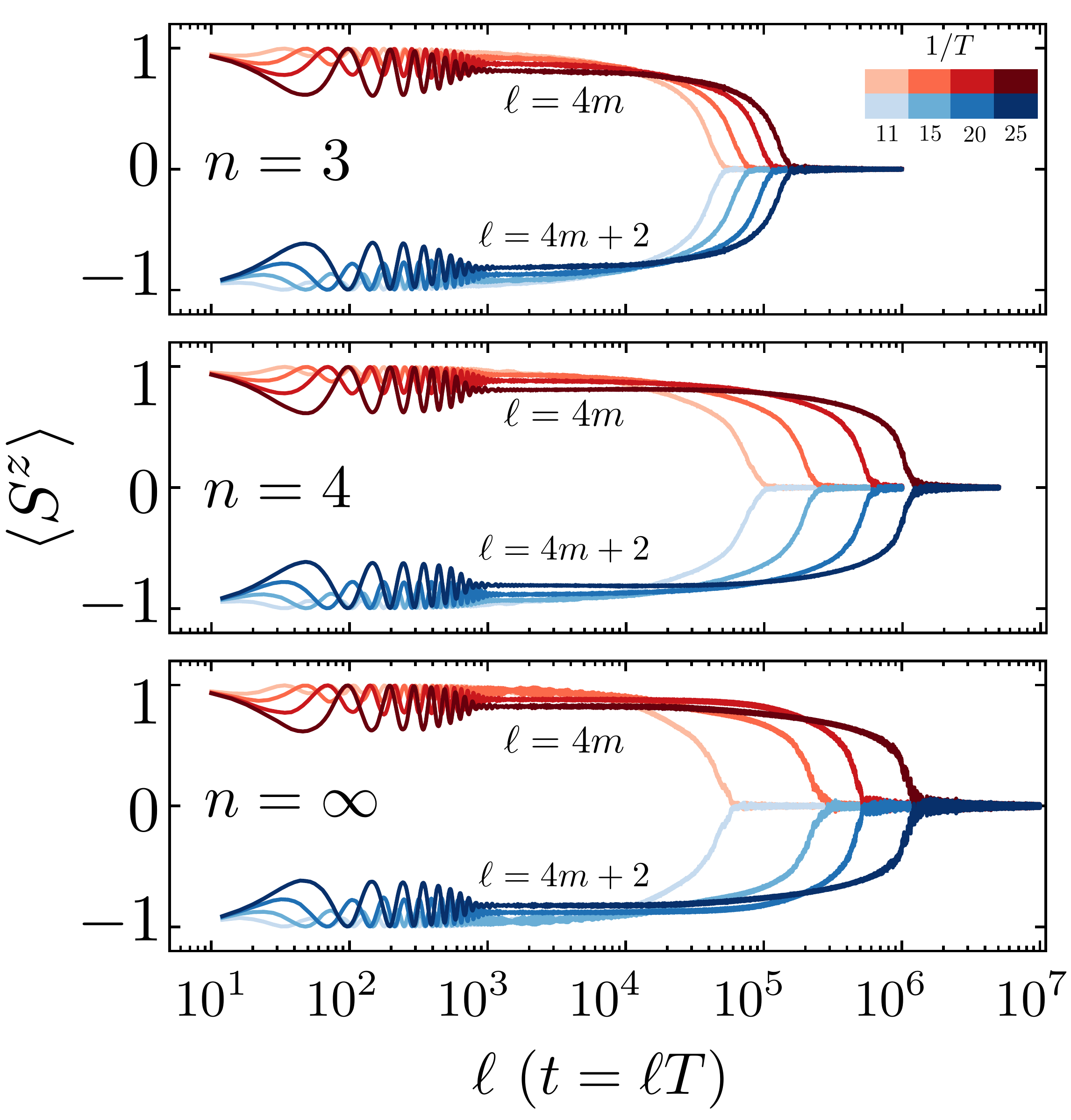}
    \caption{{\bf Classical Time Rondeau Crystals:} Time evolution of the average magnetization, $\langle S^z \rangle$, for various $n$-RMDs and the Thue-Morse drive, when the transverse field $g = 0.255 \omega$, where $\omega = 2\pi/T$. Long-lived oscillations of $\langle S^z \rangle$ are observed signalling the existence of a classical prethermal time rondeau crystal phase (see text). 
    }
    \label{fig:rondeau_time_crystals}
\end{figure}

In this section, we have presented compelling numerical evidence demonstrating that the thermalization process of classical spin models is parametrically slower than their quantum counterparts. We now discuss a conjecture for the possible reason for this longer-lived prethermalization. We start by noting that the classical model emerges in the $S \rightarrow \infty$ limit of a quantum spin-$S$ model~\cite{mori2018floquet}. Intriguingly, the dynamics of the spin-$S$ quantum model is equivalent to that of a long-range interacting spin-$1/2$ model~\cite{mori2018floquet} (for more details, see ref.~\cite{suppmat}). We conjecture that the longer-lived prethermal plateau in the classical regime originates from these effective long-range interactions. It is well known that strong long-range interacting quantum systems can exhibit much slower heating compared to their short-range counterparts~\cite{defenu2024out,kelly2021stroboscopic,pizzi2021higher}. This enhanced stability of driven long-range systems arises due to many-body scarring~\cite{lerose2023theory}. Notably, both driven long-range interacting quantum systems and short-range interacting classical systems can host non-equilibrium phases of matter that are not hosted by short-range quantum models~\cite{pizzi2021classicalPRL,pizzi2021classical,pizzi2021higher,giachetti2023fractal}. In order to substantiate our conjecture, we studied the spin-$S$ quantum model (see supplementary material for details~\cite{suppmat}). However, the $1/T^{(2n+2)}$ scaling of the thermalization time did not emerge for the $S$ values that we could investigate. 

\section{Time Rondeau Crystals} 
\label{sec:TRC}
Having established the existence of classical prethermalization for both RMDs and the Thue-Morse drive, we now proceed to investigate routes to realize non-equilibrium phases of matter in these systems. To this end, we study a protocol to realize a time rondeau crystal (TRC) - a novel phase of matter characterized by long-time periodic temporal order accompanied by short-time temporal disorder~\cite{moon2024experimental}. A TRC generalizes the notion of time-translation symmetry breaking (TTSB) to aperiodically driven systems~\cite{zhao2023temporal}. In the spin$-1/2$ version of our model, a TRC can be realized by tuning the transverse magnetic field, when $g\sim 0.25 \omega$, where $\omega = \frac{2 \pi}{T}$ for the RMD protocol discussed here. In this case, the Floquet counterpart of this model would exhibit period-doubling oscillations of the magnetization for exponentially long times, thereby signifying TTSB and a resultant prethermal time-crystal order.  

We examine the time-evolution of the stroboscopic magnetization of this system at $t=4mT$ (where $m\in \mathbb{Z^+}$) for various $n-$RMDs and find that a clear prethermal TRC phase emerges in this system, when $g\sim 0.25 \omega$ and $n\ge 3$ (see fig.~\ref{fig:rondeau_time_crystals}); the lifetime of these TRCs can be controlled very effectively by tuning the driving frequency. This protocol is the RMD generalization of the periodic driving protocol discussed in ref.~\cite{pizzi2021classicalPRL}. Notably, the TRC lifetime for the random sextapolar drive protocol ($n=4$) is very close to the Thue-Morse protocol, despite the inherent randomness in the former. Furthermore, these time crystals are reasonably robust and they can be observed in the $g_{\rm tc} \in [0.244,0.56]$ regime~\cite{suppmat}. Our results indicate that structured aperiodic driving can be effectively harnessed to realize prethermal phases of matter in classical many-body systems.\\

\section{Conclusion and Outlook} 
\label{sec:conclusion}
In this work, we have explored the dynamics of a classical spin model subjected to multipolar driving. We have found that this system exhibits a robust prethermal regime for a wide range of initial states. By studying the decorrelator, we have demonstrated that the thermalization time scales algebraically $(\sim (1/T)^{2n+2}$ for $n-$RMDs. Furthermore, in the $n \rightarrow \infty$ limit, the thermalization time scales exponentially with the driving frequency. These results demonstrate that thermalization in classical many-body systems can be much slower than their quantum counterparts, despite the absence of any Lieb-Robinson bounds. Our study raises some intriguing questions about the dynamics of driven systems in the classical ($ S \rightarrow \infty$) limit. We note that similar issues have been pointed out in the context of scrambling in classical many-body systems~\cite{bilitewski2018temperature}. Finally, we demonstrate that these aperiodically driven systems can host classical prethermal phases of matter like time rondeau crystals.

There are several avenues for future work in these systems. For instance, it would be interesting to explore the scaling of the thermalization time for long-range interacting classical and quantum spin models. Exploring the fate of the time rondeau crystal order in three-dimensional lattices is another worthwhile direction for future research. Finally, it would be interesting to extend our analysis to other aperiodic driving protocols and investigate the emergence of other non-equilibrium phases of matter, such as time quasicrystals in classical many-body systems. \\

\section*{Acknowledgements}
We thank Subhro Bhattacharjee for pointing out ref.~\cite{bilitewski2018temperature}. SC thanks DST, India for support through SERB project SRG/2023/002730 and ICTS for participating in the program - Stability of Quantum Matter in and out of Equilibrium at Various Scales (code: ICTS/SQMVS2024/01). SK has been supported by the Visiting Students Program at HRI.

\bibliographystyle{apsrev4-1}
\bibliography{ref}

\begin{thebibliography}{61}%
\makeatletter
\providecommand \@ifxundefined [1]{%
 \@ifx{#1\undefined}
}%
\providecommand \@ifnum [1]{%
 \ifnum #1\expandafter \@firstoftwo
 \else \expandafter \@secondoftwo
 \fi
}%
\providecommand \@ifx [1]{%
 \ifx #1\expandafter \@firstoftwo
 \else \expandafter \@secondoftwo
 \fi
}%
\providecommand \natexlab [1]{#1}%
\providecommand \enquote  [1]{``#1''}%
\providecommand \bibnamefont  [1]{#1}%
\providecommand \bibfnamefont [1]{#1}%
\providecommand \citenamefont [1]{#1}%
\providecommand \href@noop [0]{\@secondoftwo}%
\providecommand \href [0]{\begingroup \@sanitize@url \@href}%
\providecommand \@href[1]{\@@startlink{#1}\@@href}%
\providecommand \@@href[1]{\endgroup#1\@@endlink}%
\providecommand \@sanitize@url [0]{\catcode `\\12\catcode `\$12\catcode
  `\&12\catcode `\#12\catcode `\^12\catcode `\_12\catcode `\%12\relax}%
\providecommand \@@startlink[1]{}%
\providecommand \@@endlink[0]{}%
\providecommand \url  [0]{\begingroup\@sanitize@url \@url }%
\providecommand \@url [1]{\endgroup\@href {#1}{\urlprefix }}%
\providecommand \urlprefix  [0]{URL }%
\providecommand \Eprint [0]{\href }%
\providecommand \doibase [0]{http://dx.doi.org/}%
\providecommand \selectlanguage [0]{\@gobble}%
\providecommand \bibinfo  [0]{\@secondoftwo}%
\providecommand \bibfield  [0]{\@secondoftwo}%
\providecommand \translation [1]{[#1]}%
\providecommand \BibitemOpen [0]{}%
\providecommand \bibitemStop [0]{}%
\providecommand \bibitemNoStop [0]{.\EOS\space}%
\providecommand \EOS [0]{\spacefactor3000\relax}%
\providecommand \BibitemShut  [1]{\csname bibitem#1\endcsname}%
\let\auto@bib@innerbib\@empty
\bibitem [{\citenamefont {Bukov}\ \emph {et~al.}(2015)\citenamefont {Bukov},
  \citenamefont {D'Alessio},\ and\ \citenamefont
  {Polkovnikov}}]{bukov2015universal}%
  \BibitemOpen
  \bibfield  {author} {\bibinfo {author} {\bibfnamefont {M.}~\bibnamefont
  {Bukov}}, \bibinfo {author} {\bibfnamefont {L.}~\bibnamefont {D'Alessio}}, \
  and\ \bibinfo {author} {\bibfnamefont {A.}~\bibnamefont {Polkovnikov}},\
  }\href@noop {} {\bibfield  {journal} {\bibinfo  {journal} {Adv. Phys.}\
  }\textbf {\bibinfo {volume} {64}},\ \bibinfo {pages} {139} (\bibinfo {year}
  {2015})}\BibitemShut {NoStop}%
\bibitem [{\citenamefont {Oka}\ and\ \citenamefont
  {Kitamura}(2019)}]{oka2019floquet}%
  \BibitemOpen
  \bibfield  {author} {\bibinfo {author} {\bibfnamefont {T.}~\bibnamefont
  {Oka}}\ and\ \bibinfo {author} {\bibfnamefont {S.}~\bibnamefont {Kitamura}},\
  }\href@noop {} {\bibfield  {journal} {\bibinfo  {journal} {Annu. Rev.
  Condens. Matter Phys.}\ }\textbf {\bibinfo {volume} {10}},\ \bibinfo {pages}
  {387} (\bibinfo {year} {2019})}\BibitemShut {NoStop}%
\bibitem [{\citenamefont {Moessner}\ and\ \citenamefont
  {Sondhi}(2017)}]{moessner2017equilibration}%
  \BibitemOpen
  \bibfield  {author} {\bibinfo {author} {\bibfnamefont {R.}~\bibnamefont
  {Moessner}}\ and\ \bibinfo {author} {\bibfnamefont {S.~L.}\ \bibnamefont
  {Sondhi}},\ }\href@noop {} {\bibfield  {journal} {\bibinfo  {journal} {Nat.
  Phys.}\ }\textbf {\bibinfo {volume} {13}},\ \bibinfo {pages} {424} (\bibinfo
  {year} {2017})}\BibitemShut {NoStop}%
\bibitem [{\citenamefont {Rudner}\ and\ \citenamefont
  {Lindner}(2020)}]{rudner2020band}%
  \BibitemOpen
  \bibfield  {author} {\bibinfo {author} {\bibfnamefont {M.~S.}\ \bibnamefont
  {Rudner}}\ and\ \bibinfo {author} {\bibfnamefont {N.~H.}\ \bibnamefont
  {Lindner}},\ }\href@noop {} {\bibfield  {journal} {\bibinfo  {journal} {Nat.
  Rev. Phys.}\ }\textbf {\bibinfo {volume} {2}},\ \bibinfo {pages} {229}
  (\bibinfo {year} {2020})}\BibitemShut {NoStop}%
\bibitem [{\citenamefont {Holthaus}(2015)}]{holthaus2015floquet}%
  \BibitemOpen
  \bibfield  {author} {\bibinfo {author} {\bibfnamefont {M.}~\bibnamefont
  {Holthaus}},\ }\href@noop {} {\bibfield  {journal} {\bibinfo  {journal} {J.
  Phys. B}\ }\textbf {\bibinfo {volume} {49}},\ \bibinfo {pages} {013001}
  (\bibinfo {year} {2015})}\BibitemShut {NoStop}%
\bibitem [{\citenamefont {Eckardt}(2017)}]{eckardt2017colloquium}%
  \BibitemOpen
  \bibfield  {author} {\bibinfo {author} {\bibfnamefont {A.}~\bibnamefont
  {Eckardt}},\ }\href@noop {} {\bibfield  {journal} {\bibinfo  {journal} {Rev.
  Mod. Phys.}\ }\textbf {\bibinfo {volume} {89}},\ \bibinfo {pages} {011004}
  (\bibinfo {year} {2017})}\BibitemShut {NoStop}%
\bibitem [{\citenamefont {Haldar}\ and\ \citenamefont
  {Das}(2022)}]{haldar2022statistical}%
  \BibitemOpen
  \bibfield  {author} {\bibinfo {author} {\bibfnamefont {A.}~\bibnamefont
  {Haldar}}\ and\ \bibinfo {author} {\bibfnamefont {A.}~\bibnamefont {Das}},\
  }\href@noop {} {\bibfield  {journal} {\bibinfo  {journal} {J. Phys: Cond.
  Matt.}\ }\textbf {\bibinfo {volume} {34}},\ \bibinfo {pages} {234001}
  (\bibinfo {year} {2022})}\BibitemShut {NoStop}%
\bibitem [{\citenamefont {Sen}\ \emph {et~al.}(2021)\citenamefont {Sen},
  \citenamefont {Sen},\ and\ \citenamefont {Sengupta}}]{sen2021analytic}%
  \BibitemOpen
  \bibfield  {author} {\bibinfo {author} {\bibfnamefont {A.}~\bibnamefont
  {Sen}}, \bibinfo {author} {\bibfnamefont {D.}~\bibnamefont {Sen}}, \ and\
  \bibinfo {author} {\bibfnamefont {K.}~\bibnamefont {Sengupta}},\ }\href@noop
  {} {\bibfield  {journal} {\bibinfo  {journal} {J. Phys: Cond. Matt.}\
  }\textbf {\bibinfo {volume} {33}},\ \bibinfo {pages} {443003} (\bibinfo
  {year} {2021})}\BibitemShut {NoStop}%
\bibitem [{\citenamefont {Sacha}\ and\ \citenamefont
  {Zakrzewski}(2017)}]{sacha2018review}%
  \BibitemOpen
  \bibfield  {author} {\bibinfo {author} {\bibfnamefont {K.}~\bibnamefont
  {Sacha}}\ and\ \bibinfo {author} {\bibfnamefont {J.}~\bibnamefont
  {Zakrzewski}},\ }\href@noop {} {\bibfield  {journal} {\bibinfo  {journal}
  {Rep. Prog. Phys.}\ }\textbf {\bibinfo {volume} {81}},\ \bibinfo {pages}
  {016401} (\bibinfo {year} {2017})}\BibitemShut {NoStop}%
\bibitem [{\citenamefont {Sacha}(2020)}]{sacha2020book}%
  \BibitemOpen
  \bibfield  {author} {\bibinfo {author} {\bibfnamefont {K.}~\bibnamefont
  {Sacha}},\ }\href@noop {} {\emph {\bibinfo {title} {Time Crystals}}},\
  \bibinfo {series} {Springer Series on Atomic, Optical, and Plasma Physics},
  Vol.\ \bibinfo {volume} {114}\ (\bibinfo  {publisher} {Springer, Cham,
  Switzerland},\ \bibinfo {year} {2020})\BibitemShut {NoStop}%
\bibitem [{\citenamefont {Khemani}\ \emph {et~al.}(2019)\citenamefont
  {Khemani}, \citenamefont {Moessner},\ and\ \citenamefont
  {Sondhi}}]{khemani2019review}%
  \BibitemOpen
  \bibfield  {author} {\bibinfo {author} {\bibfnamefont {V.}~\bibnamefont
  {Khemani}}, \bibinfo {author} {\bibfnamefont {R.}~\bibnamefont {Moessner}}, \
  and\ \bibinfo {author} {\bibfnamefont {S.~L.}\ \bibnamefont {Sondhi}},\
  }\href@noop {} {\bibfield  {journal} {\bibinfo  {journal} {arXiv:1910.10745}\
  } (\bibinfo {year} {2019})}\BibitemShut {NoStop}%
\bibitem [{\citenamefont {Else}\ \emph
  {et~al.}(2020{\natexlab{a}})\citenamefont {Else}, \citenamefont {Monroe},
  \citenamefont {Nayak},\ and\ \citenamefont {Yao}}]{nayak2019review}%
  \BibitemOpen
  \bibfield  {author} {\bibinfo {author} {\bibfnamefont {D.~V.}\ \bibnamefont
  {Else}}, \bibinfo {author} {\bibfnamefont {C.}~\bibnamefont {Monroe}},
  \bibinfo {author} {\bibfnamefont {C.}~\bibnamefont {Nayak}}, \ and\ \bibinfo
  {author} {\bibfnamefont {N.~Y.}\ \bibnamefont {Yao}},\ }\href@noop {}
  {\bibfield  {journal} {\bibinfo  {journal} {Annu. Rev. Condens. Matter
  Phys.}\ }\textbf {\bibinfo {volume} {11}},\ \bibinfo {pages} {467} (\bibinfo
  {year} {2020}{\natexlab{a}})}\BibitemShut {NoStop}%
\bibitem [{\citenamefont {von Keyserlingk}\ \emph {et~al.}(2016)\citenamefont
  {von Keyserlingk}, \citenamefont {Khemani},\ and\ \citenamefont
  {Sondhi}}]{sondhi2016prb}%
  \BibitemOpen
  \bibfield  {author} {\bibinfo {author} {\bibfnamefont {C.~W.}\ \bibnamefont
  {von Keyserlingk}}, \bibinfo {author} {\bibfnamefont {V.}~\bibnamefont
  {Khemani}}, \ and\ \bibinfo {author} {\bibfnamefont {S.~L.}\ \bibnamefont
  {Sondhi}},\ }\href@noop {} {\bibfield  {journal} {\bibinfo  {journal} {Phys.
  Rev. B}\ }\textbf {\bibinfo {volume} {94}},\ \bibinfo {pages} {085112}
  (\bibinfo {year} {2016})}\BibitemShut {NoStop}%
\bibitem [{\citenamefont {Khemani}\ \emph {et~al.}(2016)\citenamefont
  {Khemani}, \citenamefont {Lazarides}, \citenamefont {Moessner},\ and\
  \citenamefont {Sondhi}}]{sondhi2016prl}%
  \BibitemOpen
  \bibfield  {author} {\bibinfo {author} {\bibfnamefont {V.}~\bibnamefont
  {Khemani}}, \bibinfo {author} {\bibfnamefont {A.}~\bibnamefont {Lazarides}},
  \bibinfo {author} {\bibfnamefont {R.}~\bibnamefont {Moessner}}, \ and\
  \bibinfo {author} {\bibfnamefont {S.~L.}\ \bibnamefont {Sondhi}},\
  }\href@noop {} {\bibfield  {journal} {\bibinfo  {journal} {Phys. Rev. Lett.}\
  }\textbf {\bibinfo {volume} {116}},\ \bibinfo {pages} {250401} (\bibinfo
  {year} {2016})}\BibitemShut {NoStop}%
\bibitem [{\citenamefont {Else}\ \emph {et~al.}(2016)\citenamefont {Else},
  \citenamefont {Bauer},\ and\ \citenamefont {Nayak}}]{nayak2016prl}%
  \BibitemOpen
  \bibfield  {author} {\bibinfo {author} {\bibfnamefont {D.~V.}\ \bibnamefont
  {Else}}, \bibinfo {author} {\bibfnamefont {B.}~\bibnamefont {Bauer}}, \ and\
  \bibinfo {author} {\bibfnamefont {C.}~\bibnamefont {Nayak}},\ }\href@noop {}
  {\bibfield  {journal} {\bibinfo  {journal} {Phys. Rev. Lett.}\ }\textbf
  {\bibinfo {volume} {117}},\ \bibinfo {pages} {090402} (\bibinfo {year}
  {2016})}\BibitemShut {NoStop}%
\bibitem [{\citenamefont {Yao}\ \emph {et~al.}(2017)\citenamefont {Yao},
  \citenamefont {Potter}, \citenamefont {Potirniche},\ and\ \citenamefont
  {Vishwanath}}]{yao2017prl}%
  \BibitemOpen
  \bibfield  {author} {\bibinfo {author} {\bibfnamefont {N.~Y.}\ \bibnamefont
  {Yao}}, \bibinfo {author} {\bibfnamefont {A.~C.}\ \bibnamefont {Potter}},
  \bibinfo {author} {\bibfnamefont {I.-D.}\ \bibnamefont {Potirniche}}, \ and\
  \bibinfo {author} {\bibfnamefont {A.}~\bibnamefont {Vishwanath}},\
  }\href@noop {} {\bibfield  {journal} {\bibinfo  {journal} {Phys. Rev. Lett.}\
  }\textbf {\bibinfo {volume} {118}},\ \bibinfo {pages} {030401} (\bibinfo
  {year} {2017})}\BibitemShut {NoStop}%
\bibitem [{\citenamefont {Zhang}\ \emph {et~al.}(2017)\citenamefont {Zhang},
  \citenamefont {Hess}, \citenamefont {Kyprianidis}, \citenamefont {Becker},
  \citenamefont {Lee}, \citenamefont {Smith}, \citenamefont {Pagano},
  \citenamefont {Potirniche}, \citenamefont {Potter}, \citenamefont
  {Vishwanath}, \citenamefont {Yao},\ and\ \citenamefont
  {Monroe}}]{monroe2017nature}%
  \BibitemOpen
  \bibfield  {author} {\bibinfo {author} {\bibfnamefont {J.}~\bibnamefont
  {Zhang}}, \bibinfo {author} {\bibfnamefont {P.}~\bibnamefont {Hess}},
  \bibinfo {author} {\bibfnamefont {A.}~\bibnamefont {Kyprianidis}}, \bibinfo
  {author} {\bibfnamefont {P.}~\bibnamefont {Becker}}, \bibinfo {author}
  {\bibfnamefont {A.}~\bibnamefont {Lee}}, \bibinfo {author} {\bibfnamefont
  {J.}~\bibnamefont {Smith}}, \bibinfo {author} {\bibfnamefont
  {G.}~\bibnamefont {Pagano}}, \bibinfo {author} {\bibfnamefont {I.-D.}\
  \bibnamefont {Potirniche}}, \bibinfo {author} {\bibfnamefont {A.~C.}\
  \bibnamefont {Potter}}, \bibinfo {author} {\bibfnamefont {A.}~\bibnamefont
  {Vishwanath}}, \bibinfo {author} {\bibfnamefont {N.~Y.}\ \bibnamefont {Yao}},
  \ and\ \bibinfo {author} {\bibfnamefont {C.}~\bibnamefont {Monroe}},\
  }\href@noop {} {\bibfield  {journal} {\bibinfo  {journal} {Nature}\ }\textbf
  {\bibinfo {volume} {543}},\ \bibinfo {pages} {217} (\bibinfo {year}
  {2017})}\BibitemShut {NoStop}%
\bibitem [{\citenamefont {D’Alessio}\ and\ \citenamefont
  {Rigol}(2014)}]{d2014long}%
  \BibitemOpen
  \bibfield  {author} {\bibinfo {author} {\bibfnamefont {L.}~\bibnamefont
  {D’Alessio}}\ and\ \bibinfo {author} {\bibfnamefont {M.}~\bibnamefont
  {Rigol}},\ }\href@noop {} {\bibfield  {journal} {\bibinfo  {journal} {Phys.
  Rev. X}\ }\textbf {\bibinfo {volume} {4}},\ \bibinfo {pages} {041048}
  (\bibinfo {year} {2014})}\BibitemShut {NoStop}%
\bibitem [{\citenamefont {Lazarides}\ \emph {et~al.}(2014)\citenamefont
  {Lazarides}, \citenamefont {Das},\ and\ \citenamefont
  {Moessner}}]{lazarides2014equilibrium}%
  \BibitemOpen
  \bibfield  {author} {\bibinfo {author} {\bibfnamefont {A.}~\bibnamefont
  {Lazarides}}, \bibinfo {author} {\bibfnamefont {A.}~\bibnamefont {Das}}, \
  and\ \bibinfo {author} {\bibfnamefont {R.}~\bibnamefont {Moessner}},\
  }\href@noop {} {\bibfield  {journal} {\bibinfo  {journal} {Phys. Rev. E}\
  }\textbf {\bibinfo {volume} {90}},\ \bibinfo {pages} {012110} (\bibinfo
  {year} {2014})}\BibitemShut {NoStop}%
\bibitem [{\citenamefont {Choudhury}\ and\ \citenamefont
  {Mueller}(2014)}]{choudhury2014stability}%
  \BibitemOpen
  \bibfield  {author} {\bibinfo {author} {\bibfnamefont {S.}~\bibnamefont
  {Choudhury}}\ and\ \bibinfo {author} {\bibfnamefont {E.~J.}\ \bibnamefont
  {Mueller}},\ }\href@noop {} {\bibfield  {journal} {\bibinfo  {journal} {Phys.
  Rev. A}\ }\textbf {\bibinfo {volume} {90}},\ \bibinfo {pages} {013621}
  (\bibinfo {year} {2014})}\BibitemShut {NoStop}%
\bibitem [{\citenamefont {Choudhury}\ and\ \citenamefont
  {Mueller}(2015)}]{choudhury2015transverse}%
  \BibitemOpen
  \bibfield  {author} {\bibinfo {author} {\bibfnamefont {S.}~\bibnamefont
  {Choudhury}}\ and\ \bibinfo {author} {\bibfnamefont {E.~J.}\ \bibnamefont
  {Mueller}},\ }\href@noop {} {\bibfield  {journal} {\bibinfo  {journal} {Phys.
  Rev. A}\ }\textbf {\bibinfo {volume} {91}},\ \bibinfo {pages} {023624}
  (\bibinfo {year} {2015})}\BibitemShut {NoStop}%
\bibitem [{\citenamefont {Bilitewski}\ and\ \citenamefont
  {Cooper}(2015{\natexlab{a}})}]{bilitewski2015scattering}%
  \BibitemOpen
  \bibfield  {author} {\bibinfo {author} {\bibfnamefont {T.}~\bibnamefont
  {Bilitewski}}\ and\ \bibinfo {author} {\bibfnamefont {N.~R.}\ \bibnamefont
  {Cooper}},\ }\href@noop {} {\bibfield  {journal} {\bibinfo  {journal} {Phys.
  Rev. A}\ }\textbf {\bibinfo {volume} {91}},\ \bibinfo {pages} {033601}
  (\bibinfo {year} {2015}{\natexlab{a}})}\BibitemShut {NoStop}%
\bibitem [{\citenamefont {Bilitewski}\ and\ \citenamefont
  {Cooper}(2015{\natexlab{b}})}]{bilitewski2015population}%
  \BibitemOpen
  \bibfield  {author} {\bibinfo {author} {\bibfnamefont {T.}~\bibnamefont
  {Bilitewski}}\ and\ \bibinfo {author} {\bibfnamefont {N.~R.}\ \bibnamefont
  {Cooper}},\ }\href@noop {} {\bibfield  {journal} {\bibinfo  {journal} {Phys.
  Rev. A}\ }\textbf {\bibinfo {volume} {91}},\ \bibinfo {pages} {063611}
  (\bibinfo {year} {2015}{\natexlab{b}})}\BibitemShut {NoStop}%
\bibitem [{\citenamefont {Abanin}\ \emph {et~al.}(2017)\citenamefont {Abanin},
  \citenamefont {De~Roeck}, \citenamefont {Ho},\ and\ \citenamefont
  {Huveneers}}]{abanin2017effective}%
  \BibitemOpen
  \bibfield  {author} {\bibinfo {author} {\bibfnamefont {D.~A.}\ \bibnamefont
  {Abanin}}, \bibinfo {author} {\bibfnamefont {W.}~\bibnamefont {De~Roeck}},
  \bibinfo {author} {\bibfnamefont {W.~W.}\ \bibnamefont {Ho}}, \ and\ \bibinfo
  {author} {\bibfnamefont {F.}~\bibnamefont {Huveneers}},\ }\href@noop {}
  {\bibfield  {journal} {\bibinfo  {journal} {Phys. Rev. B}\ }\textbf {\bibinfo
  {volume} {95}},\ \bibinfo {pages} {014112} (\bibinfo {year}
  {2017})}\BibitemShut {NoStop}%
\bibitem [{\citenamefont {Morningstar}\ \emph {et~al.}(2023)\citenamefont
  {Morningstar}, \citenamefont {Huse},\ and\ \citenamefont
  {Khemani}}]{morningstar2023universality}%
  \BibitemOpen
  \bibfield  {author} {\bibinfo {author} {\bibfnamefont {A.}~\bibnamefont
  {Morningstar}}, \bibinfo {author} {\bibfnamefont {D.~A.}\ \bibnamefont
  {Huse}}, \ and\ \bibinfo {author} {\bibfnamefont {V.}~\bibnamefont
  {Khemani}},\ }\href@noop {} {\bibfield  {journal} {\bibinfo  {journal} {Phys.
  Rev. B}\ }\textbf {\bibinfo {volume} {108}},\ \bibinfo {pages} {174303}
  (\bibinfo {year} {2023})}\BibitemShut {NoStop}%
\bibitem [{\citenamefont {Ho}\ \emph {et~al.}(2023)\citenamefont {Ho},
  \citenamefont {Mori}, \citenamefont {Abanin},\ and\ \citenamefont
  {Dalla~Torre}}]{ho2023quantum}%
  \BibitemOpen
  \bibfield  {author} {\bibinfo {author} {\bibfnamefont {W.~W.}\ \bibnamefont
  {Ho}}, \bibinfo {author} {\bibfnamefont {T.}~\bibnamefont {Mori}}, \bibinfo
  {author} {\bibfnamefont {D.~A.}\ \bibnamefont {Abanin}}, \ and\ \bibinfo
  {author} {\bibfnamefont {E.~G.}\ \bibnamefont {Dalla~Torre}},\ }\href@noop {}
  {\bibfield  {journal} {\bibinfo  {journal} {Annals of Physics}\ ,\ \bibinfo
  {pages} {169297}} (\bibinfo {year} {2023})}\BibitemShut {NoStop}%
\bibitem [{\citenamefont {Machado}\ \emph {et~al.}(2019)\citenamefont
  {Machado}, \citenamefont {Kahanamoku-Meyer}, \citenamefont {Else},
  \citenamefont {Nayak},\ and\ \citenamefont {Yao}}]{machado2019exponentially}%
  \BibitemOpen
  \bibfield  {author} {\bibinfo {author} {\bibfnamefont {F.}~\bibnamefont
  {Machado}}, \bibinfo {author} {\bibfnamefont {G.~D.}\ \bibnamefont
  {Kahanamoku-Meyer}}, \bibinfo {author} {\bibfnamefont {D.~V.}\ \bibnamefont
  {Else}}, \bibinfo {author} {\bibfnamefont {C.}~\bibnamefont {Nayak}}, \ and\
  \bibinfo {author} {\bibfnamefont {N.~Y.}\ \bibnamefont {Yao}},\ }\href@noop
  {} {\bibfield  {journal} {\bibinfo  {journal} {Phys. Rev. Research}\ }\textbf
  {\bibinfo {volume} {1}},\ \bibinfo {pages} {033202} (\bibinfo {year}
  {2019})}\BibitemShut {NoStop}%
\bibitem [{\citenamefont {Machado}\ \emph {et~al.}(2020)\citenamefont
  {Machado}, \citenamefont {Else}, \citenamefont {Kahanamoku-Meyer},
  \citenamefont {Nayak},\ and\ \citenamefont {Yao}}]{machado2020long}%
  \BibitemOpen
  \bibfield  {author} {\bibinfo {author} {\bibfnamefont {F.}~\bibnamefont
  {Machado}}, \bibinfo {author} {\bibfnamefont {D.~V.}\ \bibnamefont {Else}},
  \bibinfo {author} {\bibfnamefont {G.~D.}\ \bibnamefont {Kahanamoku-Meyer}},
  \bibinfo {author} {\bibfnamefont {C.}~\bibnamefont {Nayak}}, \ and\ \bibinfo
  {author} {\bibfnamefont {N.~Y.}\ \bibnamefont {Yao}},\ }\href@noop {}
  {\bibfield  {journal} {\bibinfo  {journal} {Phys. Rev. X}\ }\textbf {\bibinfo
  {volume} {10}},\ \bibinfo {pages} {011043} (\bibinfo {year}
  {2020})}\BibitemShut {NoStop}%
\bibitem [{\citenamefont {Else}\ \emph {et~al.}(2017)\citenamefont {Else},
  \citenamefont {Bauer},\ and\ \citenamefont {Nayak}}]{else2017prethermal}%
  \BibitemOpen
  \bibfield  {author} {\bibinfo {author} {\bibfnamefont {D.~V.}\ \bibnamefont
  {Else}}, \bibinfo {author} {\bibfnamefont {B.}~\bibnamefont {Bauer}}, \ and\
  \bibinfo {author} {\bibfnamefont {C.}~\bibnamefont {Nayak}},\ }\href@noop {}
  {\bibfield  {journal} {\bibinfo  {journal} {Phys. Rev. X}\ }\textbf {\bibinfo
  {volume} {7}},\ \bibinfo {pages} {011026} (\bibinfo {year}
  {2017})}\BibitemShut {NoStop}%
\bibitem [{\citenamefont {Kyprianidis}\ \emph {et~al.}(2021)\citenamefont
  {Kyprianidis}, \citenamefont {Machado}, \citenamefont {Morong}, \citenamefont
  {Becker}, \citenamefont {Collins}, \citenamefont {Else}, \citenamefont
  {Feng}, \citenamefont {Hess}, \citenamefont {Nayak}, \citenamefont {Pagano}
  \emph {et~al.}}]{kyprianidis2021observation}%
  \BibitemOpen
  \bibfield  {author} {\bibinfo {author} {\bibfnamefont {A.}~\bibnamefont
  {Kyprianidis}}, \bibinfo {author} {\bibfnamefont {F.}~\bibnamefont
  {Machado}}, \bibinfo {author} {\bibfnamefont {W.}~\bibnamefont {Morong}},
  \bibinfo {author} {\bibfnamefont {P.}~\bibnamefont {Becker}}, \bibinfo
  {author} {\bibfnamefont {K.~S.}\ \bibnamefont {Collins}}, \bibinfo {author}
  {\bibfnamefont {D.~V.}\ \bibnamefont {Else}}, \bibinfo {author}
  {\bibfnamefont {L.}~\bibnamefont {Feng}}, \bibinfo {author} {\bibfnamefont
  {P.~W.}\ \bibnamefont {Hess}}, \bibinfo {author} {\bibfnamefont
  {C.}~\bibnamefont {Nayak}}, \bibinfo {author} {\bibfnamefont
  {G.}~\bibnamefont {Pagano}},  \emph {et~al.},\ }\href@noop {} {\bibfield
  {journal} {\bibinfo  {journal} {Science}\ }\textbf {\bibinfo {volume}
  {372}},\ \bibinfo {pages} {1192} (\bibinfo {year} {2021})}\BibitemShut
  {NoStop}%
\bibitem [{\citenamefont {O’Dea}\ \emph {et~al.}(2024)\citenamefont
  {O’Dea}, \citenamefont {Burnell}, \citenamefont {Chandran},\ and\
  \citenamefont {Khemani}}]{o2024prethermal}%
  \BibitemOpen
  \bibfield  {author} {\bibinfo {author} {\bibfnamefont {N.}~\bibnamefont
  {O’Dea}}, \bibinfo {author} {\bibfnamefont {F.}~\bibnamefont {Burnell}},
  \bibinfo {author} {\bibfnamefont {A.}~\bibnamefont {Chandran}}, \ and\
  \bibinfo {author} {\bibfnamefont {V.}~\bibnamefont {Khemani}},\ }\href@noop
  {} {\bibfield  {journal} {\bibinfo  {journal} {Phys. Rev. Lett.}\ }\textbf
  {\bibinfo {volume} {132}},\ \bibinfo {pages} {100401} (\bibinfo {year}
  {2024})}\BibitemShut {NoStop}%
\bibitem [{\citenamefont {Dumitrescu}\ \emph {et~al.}(2018)\citenamefont
  {Dumitrescu}, \citenamefont {Vasseur},\ and\ \citenamefont
  {Potter}}]{dumitrescu2018logarithmically}%
  \BibitemOpen
  \bibfield  {author} {\bibinfo {author} {\bibfnamefont {P.~T.}\ \bibnamefont
  {Dumitrescu}}, \bibinfo {author} {\bibfnamefont {R.}~\bibnamefont {Vasseur}},
  \ and\ \bibinfo {author} {\bibfnamefont {A.~C.}\ \bibnamefont {Potter}},\
  }\href@noop {} {\bibfield  {journal} {\bibinfo  {journal} {Phys. Rev. Lett.}\
  }\textbf {\bibinfo {volume} {120}},\ \bibinfo {pages} {070602} (\bibinfo
  {year} {2018})}\BibitemShut {NoStop}%
\bibitem [{\citenamefont {Else}\ \emph
  {et~al.}(2020{\natexlab{b}})\citenamefont {Else}, \citenamefont {Ho},\ and\
  \citenamefont {Dumitrescu}}]{else2020long}%
  \BibitemOpen
  \bibfield  {author} {\bibinfo {author} {\bibfnamefont {D.~V.}\ \bibnamefont
  {Else}}, \bibinfo {author} {\bibfnamefont {W.~W.}\ \bibnamefont {Ho}}, \ and\
  \bibinfo {author} {\bibfnamefont {P.~T.}\ \bibnamefont {Dumitrescu}},\
  }\href@noop {} {\bibfield  {journal} {\bibinfo  {journal} {Phys. Rev. X}\
  }\textbf {\bibinfo {volume} {10}},\ \bibinfo {pages} {021032} (\bibinfo
  {year} {2020}{\natexlab{b}})}\BibitemShut {NoStop}%
\bibitem [{\citenamefont {Nandy}\ \emph {et~al.}(2017)\citenamefont {Nandy},
  \citenamefont {Sen},\ and\ \citenamefont {Sen}}]{nandy2017aperiodically}%
  \BibitemOpen
  \bibfield  {author} {\bibinfo {author} {\bibfnamefont {S.}~\bibnamefont
  {Nandy}}, \bibinfo {author} {\bibfnamefont {A.}~\bibnamefont {Sen}}, \ and\
  \bibinfo {author} {\bibfnamefont {D.}~\bibnamefont {Sen}},\ }\href@noop {}
  {\bibfield  {journal} {\bibinfo  {journal} {Phys. Rev. X}\ }\textbf {\bibinfo
  {volume} {7}},\ \bibinfo {pages} {031034} (\bibinfo {year}
  {2017})}\BibitemShut {NoStop}%
\bibitem [{\citenamefont {Nandy}\ \emph {et~al.}(2018)\citenamefont {Nandy},
  \citenamefont {Sen},\ and\ \citenamefont {Sen}}]{nandy2018steady}%
  \BibitemOpen
  \bibfield  {author} {\bibinfo {author} {\bibfnamefont {S.}~\bibnamefont
  {Nandy}}, \bibinfo {author} {\bibfnamefont {A.}~\bibnamefont {Sen}}, \ and\
  \bibinfo {author} {\bibfnamefont {D.}~\bibnamefont {Sen}},\ }\href@noop {}
  {\bibfield  {journal} {\bibinfo  {journal} {Phys. Rev. B}\ }\textbf {\bibinfo
  {volume} {98}},\ \bibinfo {pages} {245144} (\bibinfo {year}
  {2018})}\BibitemShut {NoStop}%
\bibitem [{\citenamefont {Choudhury}\ and\ \citenamefont
  {Liu}(2021)}]{choudhury2021self}%
  \BibitemOpen
  \bibfield  {author} {\bibinfo {author} {\bibfnamefont {S.}~\bibnamefont
  {Choudhury}}\ and\ \bibinfo {author} {\bibfnamefont {W.~V.}\ \bibnamefont
  {Liu}},\ }\href@noop {} {\bibfield  {journal} {\bibinfo  {journal}
  {arXiv:2109.05318}\ } (\bibinfo {year} {2021})}\BibitemShut {NoStop}%
\bibitem [{\citenamefont {Chen}\ \emph {et~al.}(2024)\citenamefont {Chen},
  \citenamefont {Sacramento},\ and\ \citenamefont
  {Mondaini}}]{chen2024multifractality}%
  \BibitemOpen
  \bibfield  {author} {\bibinfo {author} {\bibfnamefont {W.}~\bibnamefont
  {Chen}}, \bibinfo {author} {\bibfnamefont {P.~D.}\ \bibnamefont
  {Sacramento}}, \ and\ \bibinfo {author} {\bibfnamefont {R.}~\bibnamefont
  {Mondaini}},\ }\href@noop {} {\bibfield  {journal} {\bibinfo  {journal}
  {Phys. Rev. B}\ }\textbf {\bibinfo {volume} {109}},\ \bibinfo {pages}
  {054202} (\bibinfo {year} {2024})}\BibitemShut {NoStop}%
\bibitem [{\citenamefont {Gallone}\ and\ \citenamefont
  {Langella}(2023)}]{gallone2023prethermalization}%
  \BibitemOpen
  \bibfield  {author} {\bibinfo {author} {\bibfnamefont {M.}~\bibnamefont
  {Gallone}}\ and\ \bibinfo {author} {\bibfnamefont {B.}~\bibnamefont
  {Langella}},\ }\href@noop {} {\bibfield  {journal} {\bibinfo  {journal}
  {arXiv preprint arXiv:2306.14022}\ } (\bibinfo {year} {2023})}\BibitemShut
  {NoStop}%
\bibitem [{\citenamefont {He}\ \emph {et~al.}(2023)\citenamefont {He},
  \citenamefont {Ye}, \citenamefont {Gong}, \citenamefont {Liu}, \citenamefont
  {Murch}, \citenamefont {Yao},\ and\ \citenamefont {Zu}}]{he2023quasi}%
  \BibitemOpen
  \bibfield  {author} {\bibinfo {author} {\bibfnamefont {G.}~\bibnamefont
  {He}}, \bibinfo {author} {\bibfnamefont {B.}~\bibnamefont {Ye}}, \bibinfo
  {author} {\bibfnamefont {R.}~\bibnamefont {Gong}}, \bibinfo {author}
  {\bibfnamefont {Z.}~\bibnamefont {Liu}}, \bibinfo {author} {\bibfnamefont
  {K.~W.}\ \bibnamefont {Murch}}, \bibinfo {author} {\bibfnamefont {N.~Y.}\
  \bibnamefont {Yao}}, \ and\ \bibinfo {author} {\bibfnamefont
  {C.}~\bibnamefont {Zu}},\ }\href@noop {} {\bibfield  {journal} {\bibinfo
  {journal} {Phys. Rev. Lett.}\ }\textbf {\bibinfo {volume} {131}},\ \bibinfo
  {pages} {130401} (\bibinfo {year} {2023})}\BibitemShut {NoStop}%
\bibitem [{\citenamefont {He}\ \emph {et~al.}(2024)\citenamefont {He},
  \citenamefont {Ye}, \citenamefont {Gong}, \citenamefont {Yao}, \citenamefont
  {Liu}, \citenamefont {Murch}, \citenamefont {Yao},\ and\ \citenamefont
  {Zu}}]{he2024exp}%
  \BibitemOpen
  \bibfield  {author} {\bibinfo {author} {\bibfnamefont {G.}~\bibnamefont
  {He}}, \bibinfo {author} {\bibfnamefont {B.}~\bibnamefont {Ye}}, \bibinfo
  {author} {\bibfnamefont {R.}~\bibnamefont {Gong}}, \bibinfo {author}
  {\bibfnamefont {C.}~\bibnamefont {Yao}}, \bibinfo {author} {\bibfnamefont
  {Z.}~\bibnamefont {Liu}}, \bibinfo {author} {\bibfnamefont {K.~W.}\
  \bibnamefont {Murch}}, \bibinfo {author} {\bibfnamefont {N.~Y.}\ \bibnamefont
  {Yao}}, \ and\ \bibinfo {author} {\bibfnamefont {C.}~\bibnamefont {Zu}},\
  }\href@noop {} {\bibfield  {journal} {\bibinfo  {journal} {arXiv preprint
  arXiv:2403.17842}\ } (\bibinfo {year} {2024})}\BibitemShut {NoStop}%
\bibitem [{\citenamefont {Dumitrescu}\ \emph {et~al.}(2022)\citenamefont
  {Dumitrescu}, \citenamefont {Bohnet}, \citenamefont {Gaebler}, \citenamefont
  {Hankin}, \citenamefont {Hayes}, \citenamefont {Kumar}, \citenamefont
  {Neyenhuis}, \citenamefont {Vasseur},\ and\ \citenamefont
  {Potter}}]{dumitrescu2022dynamical}%
  \BibitemOpen
  \bibfield  {author} {\bibinfo {author} {\bibfnamefont {P.~T.}\ \bibnamefont
  {Dumitrescu}}, \bibinfo {author} {\bibfnamefont {J.~G.}\ \bibnamefont
  {Bohnet}}, \bibinfo {author} {\bibfnamefont {J.~P.}\ \bibnamefont {Gaebler}},
  \bibinfo {author} {\bibfnamefont {A.}~\bibnamefont {Hankin}}, \bibinfo
  {author} {\bibfnamefont {D.}~\bibnamefont {Hayes}}, \bibinfo {author}
  {\bibfnamefont {A.}~\bibnamefont {Kumar}}, \bibinfo {author} {\bibfnamefont
  {B.}~\bibnamefont {Neyenhuis}}, \bibinfo {author} {\bibfnamefont
  {R.}~\bibnamefont {Vasseur}}, \ and\ \bibinfo {author} {\bibfnamefont
  {A.~C.}\ \bibnamefont {Potter}},\ }\href@noop {} {\bibfield  {journal}
  {\bibinfo  {journal} {Nature}\ }\textbf {\bibinfo {volume} {607}},\ \bibinfo
  {pages} {463} (\bibinfo {year} {2022})}\BibitemShut {NoStop}%
\bibitem [{\citenamefont {Zhao}\ \emph {et~al.}(2021)\citenamefont {Zhao},
  \citenamefont {Mintert}, \citenamefont {Moessner},\ and\ \citenamefont
  {Knolle}}]{zhao2021random}%
  \BibitemOpen
  \bibfield  {author} {\bibinfo {author} {\bibfnamefont {H.}~\bibnamefont
  {Zhao}}, \bibinfo {author} {\bibfnamefont {F.}~\bibnamefont {Mintert}},
  \bibinfo {author} {\bibfnamefont {R.}~\bibnamefont {Moessner}}, \ and\
  \bibinfo {author} {\bibfnamefont {J.}~\bibnamefont {Knolle}},\ }\href@noop {}
  {\bibfield  {journal} {\bibinfo  {journal} {Phys. Rev. Lett.}\ }\textbf
  {\bibinfo {volume} {126}},\ \bibinfo {pages} {040601} (\bibinfo {year}
  {2021})}\BibitemShut {NoStop}%
\bibitem [{\citenamefont {Mori}\ \emph {et~al.}(2021)\citenamefont {Mori},
  \citenamefont {Zhao}, \citenamefont {Mintert}, \citenamefont {Knolle},\ and\
  \citenamefont {Moessner}}]{mori2021rigorous}%
  \BibitemOpen
  \bibfield  {author} {\bibinfo {author} {\bibfnamefont {T.}~\bibnamefont
  {Mori}}, \bibinfo {author} {\bibfnamefont {H.}~\bibnamefont {Zhao}}, \bibinfo
  {author} {\bibfnamefont {F.}~\bibnamefont {Mintert}}, \bibinfo {author}
  {\bibfnamefont {J.}~\bibnamefont {Knolle}}, \ and\ \bibinfo {author}
  {\bibfnamefont {R.}~\bibnamefont {Moessner}},\ }\href@noop {} {\bibfield
  {journal} {\bibinfo  {journal} {Phys. Rev. Lett.}\ }\textbf {\bibinfo
  {volume} {127}},\ \bibinfo {pages} {050602} (\bibinfo {year}
  {2021})}\BibitemShut {NoStop}%
\bibitem [{\citenamefont {Zhao}\ \emph {et~al.}(2023)\citenamefont {Zhao},
  \citenamefont {Knolle},\ and\ \citenamefont {Moessner}}]{zhao2023temporal}%
  \BibitemOpen
  \bibfield  {author} {\bibinfo {author} {\bibfnamefont {H.}~\bibnamefont
  {Zhao}}, \bibinfo {author} {\bibfnamefont {J.}~\bibnamefont {Knolle}}, \ and\
  \bibinfo {author} {\bibfnamefont {R.}~\bibnamefont {Moessner}},\ }\href@noop
  {} {\bibfield  {journal} {\bibinfo  {journal} {Phys. Rev. B}\ }\textbf
  {\bibinfo {volume} {108}},\ \bibinfo {pages} {L100203} (\bibinfo {year}
  {2023})}\BibitemShut {NoStop}%
\bibitem [{\citenamefont {Yan}\ \emph {et~al.}(2024)\citenamefont {Yan},
  \citenamefont {Moessner},\ and\ \citenamefont
  {Zhao}}]{yan2024prethermalization}%
  \BibitemOpen
  \bibfield  {author} {\bibinfo {author} {\bibfnamefont {J.}~\bibnamefont
  {Yan}}, \bibinfo {author} {\bibfnamefont {R.}~\bibnamefont {Moessner}}, \
  and\ \bibinfo {author} {\bibfnamefont {H.}~\bibnamefont {Zhao}},\ }\href@noop
  {} {\bibfield  {journal} {\bibinfo  {journal} {Phys. Rev. B}\ }\textbf
  {\bibinfo {volume} {109}},\ \bibinfo {pages} {064305} (\bibinfo {year}
  {2024})}\BibitemShut {NoStop}%
\bibitem [{\citenamefont {Moon}\ \emph {et~al.}(2024)\citenamefont {Moon},
  \citenamefont {Schindler}, \citenamefont {Sun}, \citenamefont {Druga},
  \citenamefont {Knolle}, \citenamefont {Moessner}, \citenamefont {Zhao},
  \citenamefont {Bukov},\ and\ \citenamefont {Ajoy}}]{moon2024experimental}%
  \BibitemOpen
  \bibfield  {author} {\bibinfo {author} {\bibfnamefont {L.~J.~I.}\
  \bibnamefont {Moon}}, \bibinfo {author} {\bibfnamefont {P.~M.}\ \bibnamefont
  {Schindler}}, \bibinfo {author} {\bibfnamefont {Y.}~\bibnamefont {Sun}},
  \bibinfo {author} {\bibfnamefont {E.}~\bibnamefont {Druga}}, \bibinfo
  {author} {\bibfnamefont {J.}~\bibnamefont {Knolle}}, \bibinfo {author}
  {\bibfnamefont {R.}~\bibnamefont {Moessner}}, \bibinfo {author}
  {\bibfnamefont {H.}~\bibnamefont {Zhao}}, \bibinfo {author} {\bibfnamefont
  {M.}~\bibnamefont {Bukov}}, \ and\ \bibinfo {author} {\bibfnamefont
  {A.}~\bibnamefont {Ajoy}},\ }\href@noop {} {\bibfield  {journal} {\bibinfo
  {journal} {arXiv preprint arXiv:2404.05620}\ } (\bibinfo {year}
  {2024})}\BibitemShut {NoStop}%
\bibitem [{\citenamefont {Mori}(2018)}]{mori2018floquet}%
  \BibitemOpen
  \bibfield  {author} {\bibinfo {author} {\bibfnamefont {T.}~\bibnamefont
  {Mori}},\ }\href@noop {} {\bibfield  {journal} {\bibinfo  {journal} {Phys.
  Rev. B}\ }\textbf {\bibinfo {volume} {98}},\ \bibinfo {pages} {104303}
  (\bibinfo {year} {2018})}\BibitemShut {NoStop}%
\bibitem [{\citenamefont {Howell}\ \emph {et~al.}(2019)\citenamefont {Howell},
  \citenamefont {Weinberg}, \citenamefont {Sels}, \citenamefont {Polkovnikov},\
  and\ \citenamefont {Bukov}}]{howell2019asymptotic}%
  \BibitemOpen
  \bibfield  {author} {\bibinfo {author} {\bibfnamefont {O.}~\bibnamefont
  {Howell}}, \bibinfo {author} {\bibfnamefont {P.}~\bibnamefont {Weinberg}},
  \bibinfo {author} {\bibfnamefont {D.}~\bibnamefont {Sels}}, \bibinfo {author}
  {\bibfnamefont {A.}~\bibnamefont {Polkovnikov}}, \ and\ \bibinfo {author}
  {\bibfnamefont {M.}~\bibnamefont {Bukov}},\ }\href@noop {} {\bibfield
  {journal} {\bibinfo  {journal} {Phys. Rev. Lett.}\ }\textbf {\bibinfo
  {volume} {122}},\ \bibinfo {pages} {010602} (\bibinfo {year}
  {2019})}\BibitemShut {NoStop}%
\bibitem [{\citenamefont {Pizzi}\ \emph
  {et~al.}(2021{\natexlab{a}})\citenamefont {Pizzi}, \citenamefont
  {Nunnenkamp},\ and\ \citenamefont {Knolle}}]{pizzi2021classicalPRL}%
  \BibitemOpen
  \bibfield  {author} {\bibinfo {author} {\bibfnamefont {A.}~\bibnamefont
  {Pizzi}}, \bibinfo {author} {\bibfnamefont {A.}~\bibnamefont {Nunnenkamp}}, \
  and\ \bibinfo {author} {\bibfnamefont {J.}~\bibnamefont {Knolle}},\
  }\href@noop {} {\bibfield  {journal} {\bibinfo  {journal} {Phys. Rev. Lett.}\
  }\textbf {\bibinfo {volume} {127}},\ \bibinfo {pages} {140602} (\bibinfo
  {year} {2021}{\natexlab{a}})}\BibitemShut {NoStop}%
\bibitem [{\citenamefont {Pizzi}\ \emph
  {et~al.}(2021{\natexlab{b}})\citenamefont {Pizzi}, \citenamefont
  {Nunnenkamp},\ and\ \citenamefont {Knolle}}]{pizzi2021classical}%
  \BibitemOpen
  \bibfield  {author} {\bibinfo {author} {\bibfnamefont {A.}~\bibnamefont
  {Pizzi}}, \bibinfo {author} {\bibfnamefont {A.}~\bibnamefont {Nunnenkamp}}, \
  and\ \bibinfo {author} {\bibfnamefont {J.}~\bibnamefont {Knolle}},\
  }\href@noop {} {\bibfield  {journal} {\bibinfo  {journal} {Phys. Rev. B}\
  }\textbf {\bibinfo {volume} {104}},\ \bibinfo {pages} {094308} (\bibinfo
  {year} {2021}{\natexlab{b}})}\BibitemShut {NoStop}%
\bibitem [{\citenamefont {Ye}\ \emph {et~al.}(2021)\citenamefont {Ye},
  \citenamefont {Machado},\ and\ \citenamefont {Yao}}]{ye2021floquet}%
  \BibitemOpen
  \bibfield  {author} {\bibinfo {author} {\bibfnamefont {B.}~\bibnamefont
  {Ye}}, \bibinfo {author} {\bibfnamefont {F.}~\bibnamefont {Machado}}, \ and\
  \bibinfo {author} {\bibfnamefont {N.~Y.}\ \bibnamefont {Yao}},\ }\href@noop
  {} {\bibfield  {journal} {\bibinfo  {journal} {Phys. Rev. Lett.}\ }\textbf
  {\bibinfo {volume} {127}},\ \bibinfo {pages} {140603} (\bibinfo {year}
  {2021})}\BibitemShut {NoStop}%
\bibitem [{sup()}]{suppmat}%
  \BibitemOpen
  \href@noop {} {}\bibinfo {note} {Supplementary Material}\BibitemShut
  {NoStop}%
\bibitem [{\citenamefont {Das}\ \emph {et~al.}(2018)\citenamefont {Das},
  \citenamefont {Chakrabarty}, \citenamefont {Dhar}, \citenamefont {Kundu},
  \citenamefont {Huse}, \citenamefont {Moessner}, \citenamefont {Ray},\ and\
  \citenamefont {Bhattacharjee}}]{das2018light}%
  \BibitemOpen
  \bibfield  {author} {\bibinfo {author} {\bibfnamefont {A.}~\bibnamefont
  {Das}}, \bibinfo {author} {\bibfnamefont {S.}~\bibnamefont {Chakrabarty}},
  \bibinfo {author} {\bibfnamefont {A.}~\bibnamefont {Dhar}}, \bibinfo {author}
  {\bibfnamefont {A.}~\bibnamefont {Kundu}}, \bibinfo {author} {\bibfnamefont
  {D.~A.}\ \bibnamefont {Huse}}, \bibinfo {author} {\bibfnamefont
  {R.}~\bibnamefont {Moessner}}, \bibinfo {author} {\bibfnamefont {S.~S.}\
  \bibnamefont {Ray}}, \ and\ \bibinfo {author} {\bibfnamefont
  {S.}~\bibnamefont {Bhattacharjee}},\ }\href@noop {} {\bibfield  {journal}
  {\bibinfo  {journal} {Phys. Rev. Lett.}\ }\textbf {\bibinfo {volume} {121}},\
  \bibinfo {pages} {024101} (\bibinfo {year} {2018})}\BibitemShut {NoStop}%
\bibitem [{\citenamefont {Bilitewski}\ \emph {et~al.}(2018)\citenamefont
  {Bilitewski}, \citenamefont {Bhattacharjee},\ and\ \citenamefont
  {Moessner}}]{bilitewski2018temperature}%
  \BibitemOpen
  \bibfield  {author} {\bibinfo {author} {\bibfnamefont {T.}~\bibnamefont
  {Bilitewski}}, \bibinfo {author} {\bibfnamefont {S.}~\bibnamefont
  {Bhattacharjee}}, \ and\ \bibinfo {author} {\bibfnamefont {R.}~\bibnamefont
  {Moessner}},\ }\href@noop {} {\bibfield  {journal} {\bibinfo  {journal}
  {Phys. Rev. Lett.}\ }\textbf {\bibinfo {volume} {121}},\ \bibinfo {pages}
  {250602} (\bibinfo {year} {2018})}\BibitemShut {NoStop}%
\bibitem [{\citenamefont {Bilitewski}\ \emph {et~al.}(2021)\citenamefont
  {Bilitewski}, \citenamefont {Bhattacharjee},\ and\ \citenamefont
  {Moessner}}]{bilitewski2021classical}%
  \BibitemOpen
  \bibfield  {author} {\bibinfo {author} {\bibfnamefont {T.}~\bibnamefont
  {Bilitewski}}, \bibinfo {author} {\bibfnamefont {S.}~\bibnamefont
  {Bhattacharjee}}, \ and\ \bibinfo {author} {\bibfnamefont {R.}~\bibnamefont
  {Moessner}},\ }\href@noop {} {\bibfield  {journal} {\bibinfo  {journal}
  {Phys. Rev. B}\ }\textbf {\bibinfo {volume} {103}},\ \bibinfo {pages}
  {174302} (\bibinfo {year} {2021})}\BibitemShut {NoStop}%
\bibitem [{\citenamefont {Jin}\ and\ \citenamefont
  {Knolle}(2024)}]{jin2024floquet}%
  \BibitemOpen
  \bibfield  {author} {\bibinfo {author} {\bibfnamefont {H.-K.}\ \bibnamefont
  {Jin}}\ and\ \bibinfo {author} {\bibfnamefont {J.}~\bibnamefont {Knolle}},\
  }\href@noop {} {\bibfield  {journal} {\bibinfo  {journal} {arXiv preprint
  arXiv:2403.17118}\ } (\bibinfo {year} {2024})}\BibitemShut {NoStop}%
\bibitem [{\citenamefont {Defenu}\ \emph {et~al.}(2024)\citenamefont {Defenu},
  \citenamefont {Lerose},\ and\ \citenamefont {Pappalardi}}]{defenu2024out}%
  \BibitemOpen
  \bibfield  {author} {\bibinfo {author} {\bibfnamefont {N.}~\bibnamefont
  {Defenu}}, \bibinfo {author} {\bibfnamefont {A.}~\bibnamefont {Lerose}}, \
  and\ \bibinfo {author} {\bibfnamefont {S.}~\bibnamefont {Pappalardi}},\
  }\href@noop {} {\bibfield  {journal} {\bibinfo  {journal} {Physics Reports}\
  }\textbf {\bibinfo {volume} {1074}},\ \bibinfo {pages} {1} (\bibinfo {year}
  {2024})}\BibitemShut {NoStop}%
\bibitem [{\citenamefont {Kelly}\ \emph {et~al.}(2021)\citenamefont {Kelly},
  \citenamefont {Timmermans}, \citenamefont {Marino},\ and\ \citenamefont
  {Tsai}}]{kelly2021stroboscopic}%
  \BibitemOpen
  \bibfield  {author} {\bibinfo {author} {\bibfnamefont {S.~P.}\ \bibnamefont
  {Kelly}}, \bibinfo {author} {\bibfnamefont {E.}~\bibnamefont {Timmermans}},
  \bibinfo {author} {\bibfnamefont {J.}~\bibnamefont {Marino}}, \ and\ \bibinfo
  {author} {\bibfnamefont {S.-W.}\ \bibnamefont {Tsai}},\ }\href@noop {}
  {\bibfield  {journal} {\bibinfo  {journal} {SciPost Physics Core}\ }\textbf
  {\bibinfo {volume} {4}},\ \bibinfo {pages} {021} (\bibinfo {year}
  {2021})}\BibitemShut {NoStop}%
\bibitem [{\citenamefont {Pizzi}\ \emph
  {et~al.}(2021{\natexlab{c}})\citenamefont {Pizzi}, \citenamefont {Knolle},\
  and\ \citenamefont {Nunnenkamp}}]{pizzi2021higher}%
  \BibitemOpen
  \bibfield  {author} {\bibinfo {author} {\bibfnamefont {A.}~\bibnamefont
  {Pizzi}}, \bibinfo {author} {\bibfnamefont {J.}~\bibnamefont {Knolle}}, \
  and\ \bibinfo {author} {\bibfnamefont {A.}~\bibnamefont {Nunnenkamp}},\
  }\href@noop {} {\bibfield  {journal} {\bibinfo  {journal} {Nature
  communications}\ }\textbf {\bibinfo {volume} {12}},\ \bibinfo {pages} {2341}
  (\bibinfo {year} {2021}{\natexlab{c}})}\BibitemShut {NoStop}%
\bibitem [{\citenamefont {Lerose}\ \emph {et~al.}(2023)\citenamefont {Lerose},
  \citenamefont {Parolini}, \citenamefont {Fazio}, \citenamefont {Abanin},\
  and\ \citenamefont {Pappalardi}}]{lerose2023theory}%
  \BibitemOpen
  \bibfield  {author} {\bibinfo {author} {\bibfnamefont {A.}~\bibnamefont
  {Lerose}}, \bibinfo {author} {\bibfnamefont {T.}~\bibnamefont {Parolini}},
  \bibinfo {author} {\bibfnamefont {R.}~\bibnamefont {Fazio}}, \bibinfo
  {author} {\bibfnamefont {D.~A.}\ \bibnamefont {Abanin}}, \ and\ \bibinfo
  {author} {\bibfnamefont {S.}~\bibnamefont {Pappalardi}},\ }\href@noop {}
  {\bibfield  {journal} {\bibinfo  {journal} {arXiv preprint arXiv:2309.12504}\
  } (\bibinfo {year} {2023})}\BibitemShut {NoStop}%
\bibitem [{\citenamefont {Giachetti}\ \emph {et~al.}(2023)\citenamefont
  {Giachetti}, \citenamefont {Solfanelli}, \citenamefont {Correale},\ and\
  \citenamefont {Defenu}}]{giachetti2023fractal}%
  \BibitemOpen
  \bibfield  {author} {\bibinfo {author} {\bibfnamefont {G.}~\bibnamefont
  {Giachetti}}, \bibinfo {author} {\bibfnamefont {A.}~\bibnamefont
  {Solfanelli}}, \bibinfo {author} {\bibfnamefont {L.}~\bibnamefont
  {Correale}}, \ and\ \bibinfo {author} {\bibfnamefont {N.}~\bibnamefont
  {Defenu}},\ }\href@noop {} {\bibfield  {journal} {\bibinfo  {journal}
  {Physical Review B}\ }\textbf {\bibinfo {volume} {108}},\ \bibinfo {pages}
  {L140102} (\bibinfo {year} {2023})}\BibitemShut {NoStop}%
\end{thebibliography}%


\begin{thebibliography}{4}%
\makeatletter
\providecommand \@ifxundefined [1]{%
 \@ifx{#1\undefined}
}%
\providecommand \@ifnum [1]{%
 \ifnum #1\expandafter \@firstoftwo
 \else \expandafter \@secondoftwo
 \fi
}%
\providecommand \@ifx [1]{%
 \ifx #1\expandafter \@firstoftwo
 \else \expandafter \@secondoftwo
 \fi
}%
\providecommand \natexlab [1]{#1}%
\providecommand \enquote  [1]{``#1''}%
\providecommand \bibnamefont  [1]{#1}%
\providecommand \bibfnamefont [1]{#1}%
\providecommand \citenamefont [1]{#1}%
\providecommand \href@noop [0]{\@secondoftwo}%
\providecommand \href [0]{\begingroup \@sanitize@url \@href}%
\providecommand \@href[1]{\@@startlink{#1}\@@href}%
\providecommand \@@href[1]{\endgroup#1\@@endlink}%
\providecommand \@sanitize@url [0]{\catcode `\\12\catcode `\$12\catcode
  `\&12\catcode `\#12\catcode `\^12\catcode `\_12\catcode `\%12\relax}%
\providecommand \@@startlink[1]{}%
\providecommand \@@endlink[0]{}%
\providecommand \url  [0]{\begingroup\@sanitize@url \@url }%
\providecommand \@url [1]{\endgroup\@href {#1}{\urlprefix }}%
\providecommand \urlprefix  [0]{URL }%
\providecommand \Eprint [0]{\href }%
\providecommand \doibase [0]{https://doi.org/}%
\providecommand \selectlanguage [0]{\@gobble}%
\providecommand \bibinfo  [0]{\@secondoftwo}%
\providecommand \bibfield  [0]{\@secondoftwo}%
\providecommand \translation [1]{[#1]}%
\providecommand \BibitemOpen [0]{}%
\providecommand \bibitemStop [0]{}%
\providecommand \bibitemNoStop [0]{.\EOS\space}%
\providecommand \EOS [0]{\spacefactor3000\relax}%
\providecommand \BibitemShut  [1]{\csname bibitem#1\endcsname}%
\let\auto@bib@innerbib\@empty
\bibitem [{\citenamefont {Zhao}\ \emph {et~al.}(2023)\citenamefont {Zhao},
  \citenamefont {Knolle},\ and\ \citenamefont {Moessner}}]{zhao2023temporal}%
  \BibitemOpen
  \bibfield  {author} {\bibinfo {author} {\bibfnamefont {H.}~\bibnamefont
  {Zhao}}, \bibinfo {author} {\bibfnamefont {J.}~\bibnamefont {Knolle}},\ and\
  \bibinfo {author} {\bibfnamefont {R.}~\bibnamefont {Moessner}},\ }\href@noop
  {} {\bibfield  {journal} {\bibinfo  {journal} {Phys. Rev. B}\ }\textbf
  {\bibinfo {volume} {108}},\ \bibinfo {pages} {L100203} (\bibinfo {year}
  {2023})}\BibitemShut {NoStop}%
\bibitem [{\citenamefont {Moon}\ \emph {et~al.}(2024)\citenamefont {Moon},
  \citenamefont {Schindler}, \citenamefont {Sun}, \citenamefont {Druga},
  \citenamefont {Knolle}, \citenamefont {Moessner}, \citenamefont {Zhao},
  \citenamefont {Bukov},\ and\ \citenamefont {Ajoy}}]{moon2024experimental}%
  \BibitemOpen
  \bibfield  {author} {\bibinfo {author} {\bibfnamefont {L.~J.~I.}\
  \bibnamefont {Moon}}, \bibinfo {author} {\bibfnamefont {P.~M.}\ \bibnamefont
  {Schindler}}, \bibinfo {author} {\bibfnamefont {Y.}~\bibnamefont {Sun}},
  \bibinfo {author} {\bibfnamefont {E.}~\bibnamefont {Druga}}, \bibinfo
  {author} {\bibfnamefont {J.}~\bibnamefont {Knolle}}, \bibinfo {author}
  {\bibfnamefont {R.}~\bibnamefont {Moessner}}, \bibinfo {author}
  {\bibfnamefont {H.}~\bibnamefont {Zhao}}, \bibinfo {author} {\bibfnamefont
  {M.}~\bibnamefont {Bukov}},\ and\ \bibinfo {author} {\bibfnamefont
  {A.}~\bibnamefont {Ajoy}},\ }\href@noop {} {\bibfield  {journal} {\bibinfo
  {journal} {arXiv preprint arXiv:2404.05620}\ } (\bibinfo {year}
  {2024})}\BibitemShut {NoStop}%
\bibitem [{\citenamefont {Howell}\ \emph {et~al.}(2019)\citenamefont {Howell},
  \citenamefont {Weinberg}, \citenamefont {Sels}, \citenamefont {Polkovnikov},\
  and\ \citenamefont {Bukov}}]{howell2019asymptotic}%
  \BibitemOpen
  \bibfield  {author} {\bibinfo {author} {\bibfnamefont {O.}~\bibnamefont
  {Howell}}, \bibinfo {author} {\bibfnamefont {P.}~\bibnamefont {Weinberg}},
  \bibinfo {author} {\bibfnamefont {D.}~\bibnamefont {Sels}}, \bibinfo {author}
  {\bibfnamefont {A.}~\bibnamefont {Polkovnikov}},\ and\ \bibinfo {author}
  {\bibfnamefont {M.}~\bibnamefont {Bukov}},\ }\href@noop {} {\bibfield
  {journal} {\bibinfo  {journal} {Phys. Rev. Lett.}\ }\textbf {\bibinfo
  {volume} {122}},\ \bibinfo {pages} {010602} (\bibinfo {year}
  {2019})}\BibitemShut {NoStop}%
\bibitem [{\citenamefont {Mori}(2018)}]{mori2018floquet}%
  \BibitemOpen
  \bibfield  {author} {\bibinfo {author} {\bibfnamefont {T.}~\bibnamefont
  {Mori}},\ }\href@noop {} {\bibfield  {journal} {\bibinfo  {journal} {Phys.
  Rev. B}\ }\textbf {\bibinfo {volume} {98}},\ \bibinfo {pages} {104303}
  (\bibinfo {year} {2018})}\BibitemShut {NoStop}%
\end{thebibliography}%

\end{document}


\title{Supplemental Material - Prethermalization in aperiodically driven classical spin systems}
\author{Sajag Kumar}
\email{sajag.kumar@niser.ac.in}
\affiliation{School of Physical Sciences, National Institute of Science Education and Research, a CI of Homi Bhabha National Institute, Jatni 752050, India}
\author{Sayan Choudhury}
\email{sayanchoudhury@hri.res.in}
\affiliation{Harish-Chandra Research Institute, a CI of Homi Bhabha National Institute, Chhatnag Road, Jhunsi, Allahabad 211019}

\maketitle
The supplemental material covers several numerical results to provide further evidence for the central results of the main text. 

\section{The threshold value for thermalization time}

\begin{figure*}[h]
    \centering
    \includegraphics[width=\textwidth]{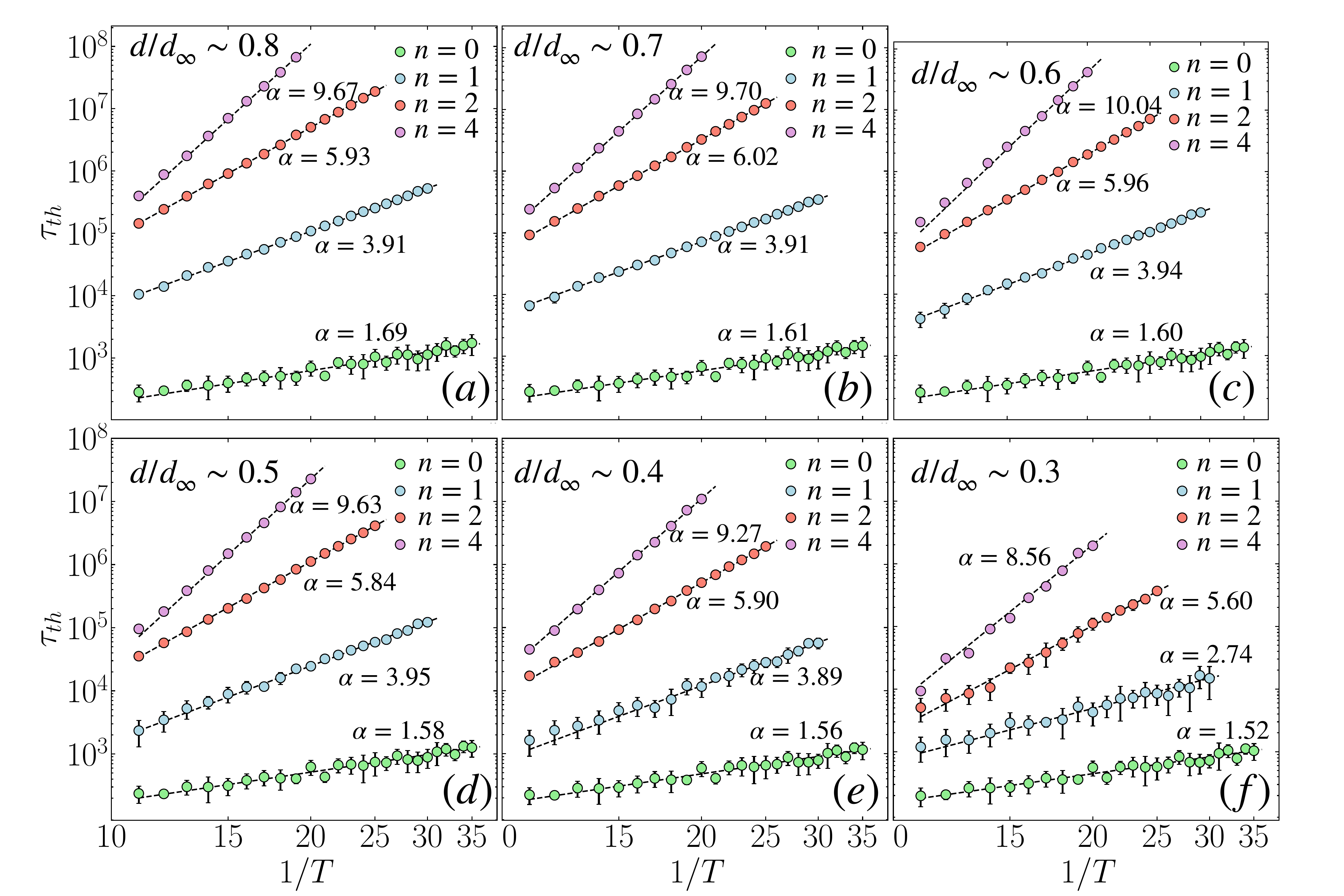}
    \caption{The scaling exponent $\alpha$ obtained by setting different values of the threshold. The threshold is denoted at the top right of each subplot. When the threshold is $d/d_{\infty} \sim x$, we extract $\tau_{th}$ by averaging over $\ell$'s for which $d(\ell T)/d_{\infty} \sim x,\, x-0.01,\, x-0.02$. It is clearly seen that for (a)-(e) $\alpha\sim 2n+2$. The usual scaling of $\alpha$ with $n$ is not observed in (f), since the threshold value reaches closer to the value of the prethermal plateau.}
    \label{fig:threshold}
\end{figure*}

We extract the thermalization time $\tau_{th}$ by setting a threshold on $d(t)/d_{\infty}$. In our analysis we have set this threshold to be $d/d_{\infty}\sim 0.9$ (We extract $\tau_{th}$ by averaging over $\ell$'s for which $d(\ell T)/d_{\infty} \sim 0.9,\, 0.89,\, 0.88$). In fig.~\ref{fig:threshold}, we show that the scaling exponent for the RMD protocol has the same $2n + 2$ behaviour for different values of the threshold (as long as the threshold is chosen away from the prethermal plateau). 

\section{The phase diagrams for the RMD scaling exponents and the Thue-Morse thermalization times}

\begin{figure*}[htp]
    \centering
    \includegraphics[width=0.7\textwidth]{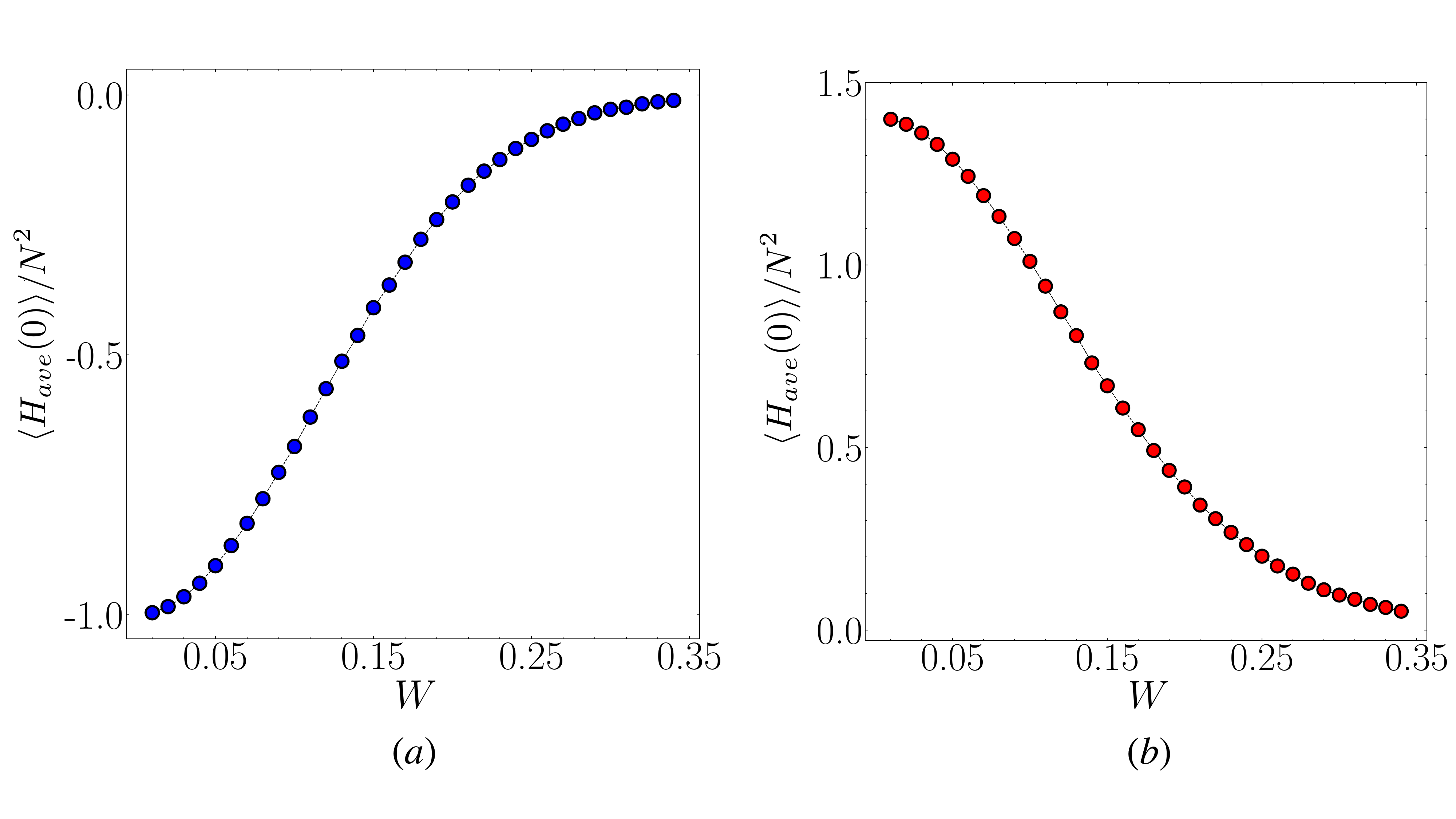}
    \caption{The dependence of the energy density of the system on the parameter $W$, (a) when the spins are anti-aligned and (b) when the spins are polarised along z-direction. To obtain each point on the plot, we average over fifty realizations of the $50 \cross 50$ system.}
    \label{fig:energy_w_plot}
\end{figure*}

\begin{figure*}[htp]
    \centering
    \includegraphics[width=\textwidth]{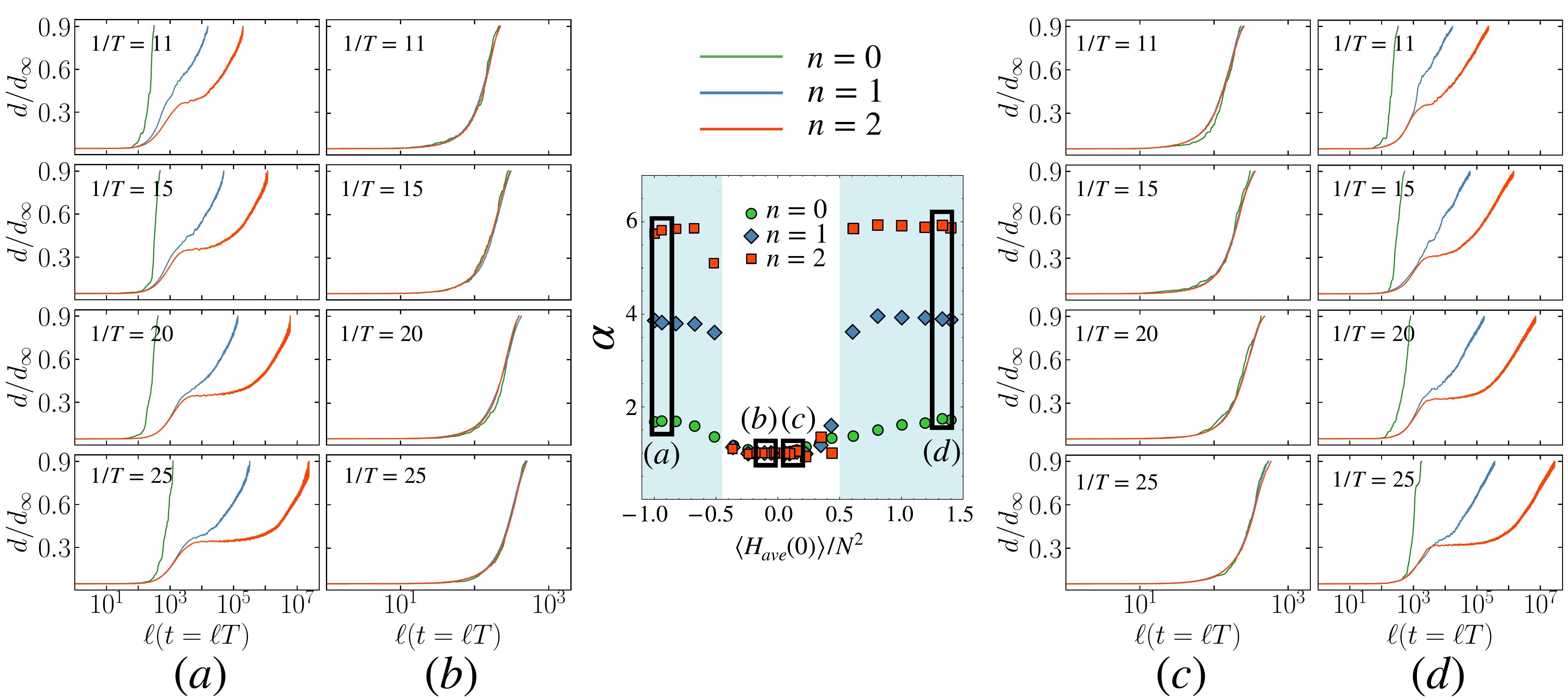}
    \caption{Some representative decorrelator evolution plots at different frequencies and for different RMD protocols from various regimes of the scaling exponent and initial energy-density phase diagram. (a) When the initial state energy density is low, we can clearly see the separation between the thermalization times of the different $n$-RMDs. In this regime the $\alpha \sim 2n+2$ scaling holds. (b) \& (c) Here, the scaling exponent becomes a constant for different $n$-RMDs (d) The initial state energy density is high, and the separation of the thermalization times for different $n$-RMDs is evident. The usual scaling $\alpha\sim 2n+2$ holds.}
    \label{fig:rmd_reps}
\end{figure*}

\begin{figure*}[htp]
    \centering
    \includegraphics[width=0.8\textwidth]{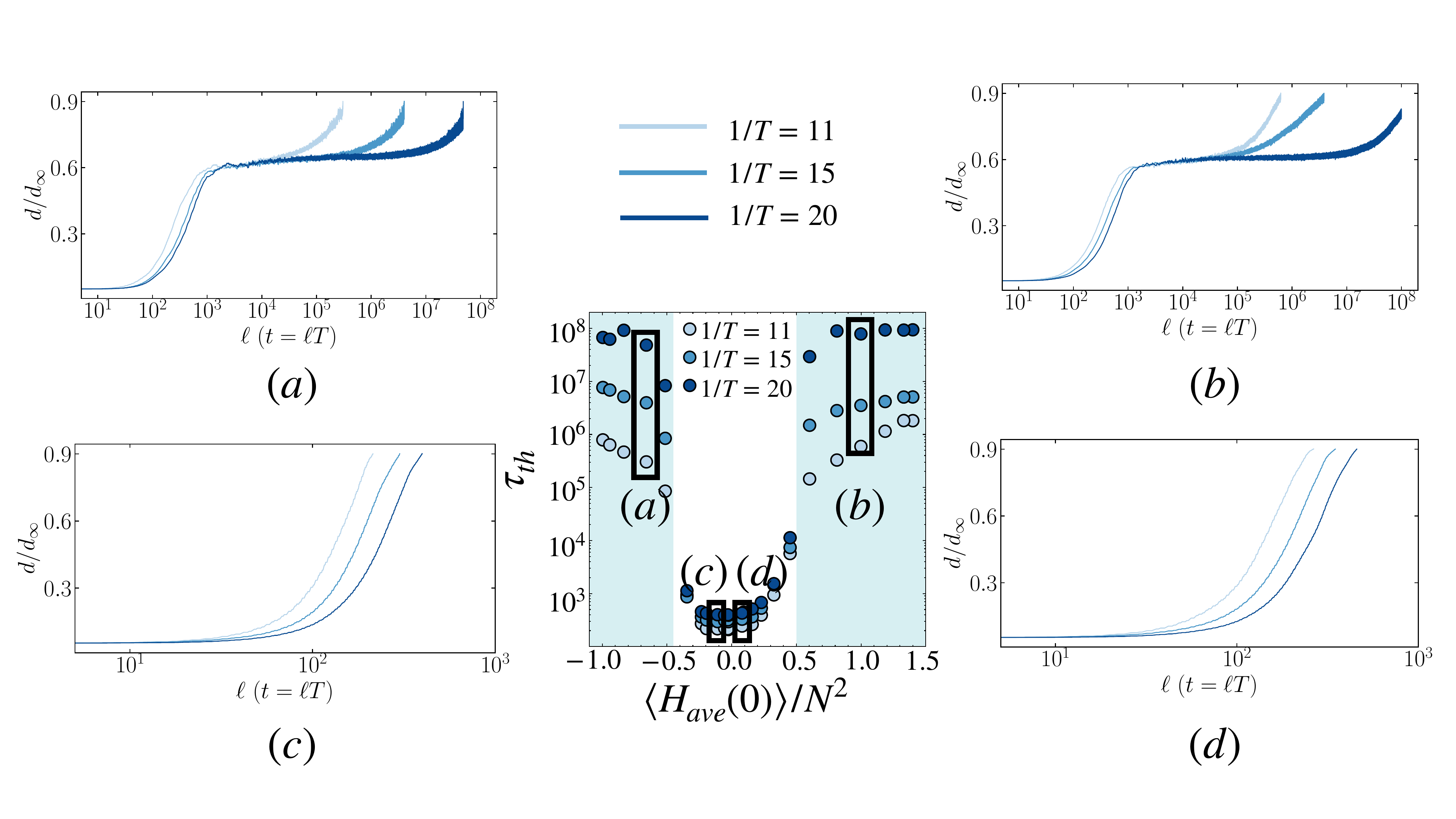}
    \caption{Some representative decorrelator evolution plots from various regimes of the thermalization time and initial energy density phase diagram for TM driving protocol. (a) \& (b) Show the decorrelator evolution for systems initialised in very low and high energy density states, respectively. The thermalization time scales exponentially in this regime. (c) \& (d) Show the decorrelator evolution for systems initialised in states with energy density close to zero. 
    }
     \label{fig:tm_reps}
\end{figure*}

In fig.~\ref{fig:energy_w_plot}, we show the dependence of the initial state that we prepare for our simulations on the parameter $W$. We initialized the system in different initial states for computing the phase diagrams by varying this parameter. In figs.~\ref{fig:rmd_reps} and \ref{fig:tm_reps}, we show some representative decorrelator evolution plots from different regimes of the phase diagrams. 

\newpage

\section{Zero longitudinal field}

\begin{figure*}[ht!]
    \centering
    \includegraphics[width=0.75\textwidth]{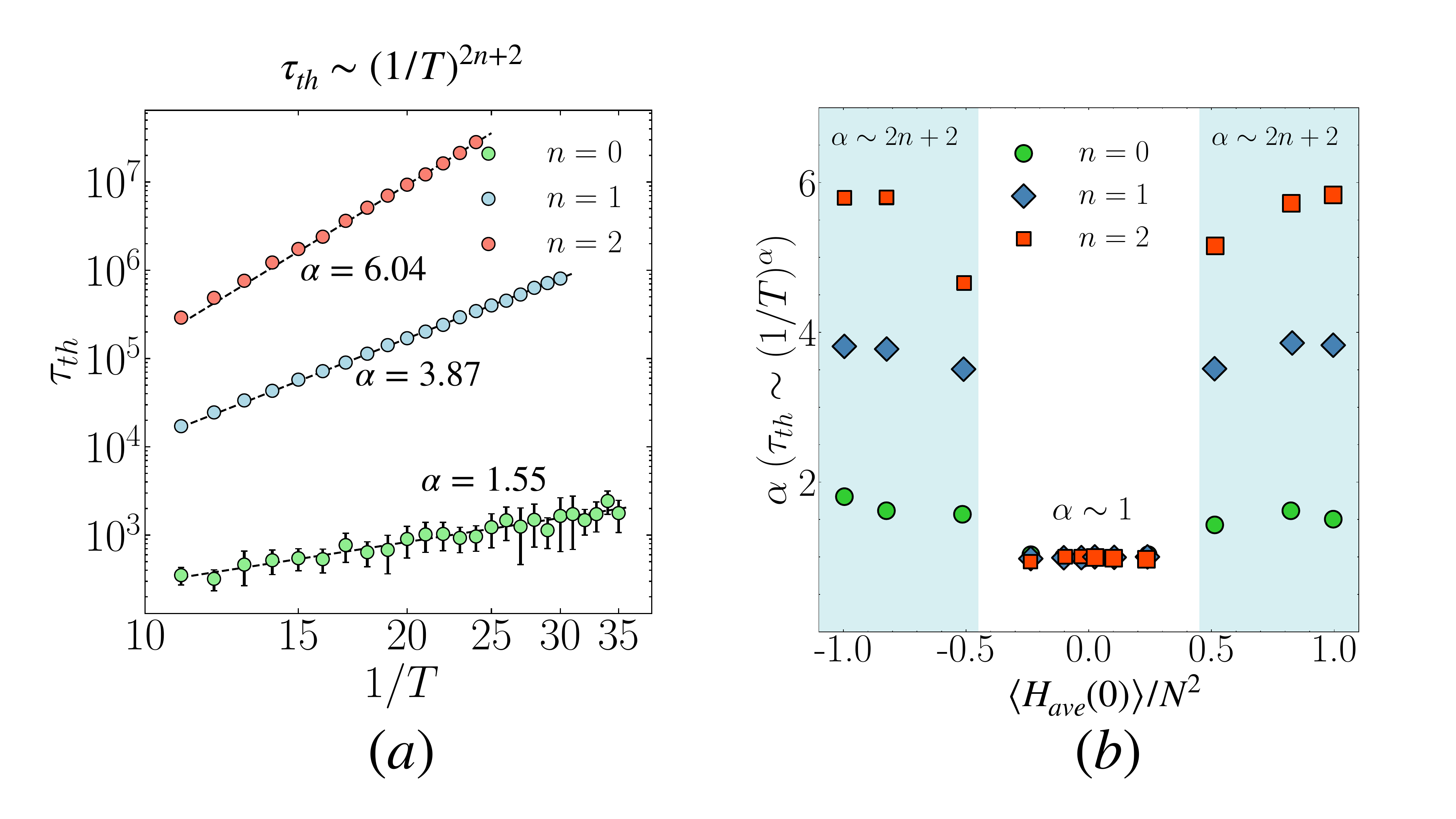}
    \caption{(a) The scaling of the thermalization time with frequency for the RMD protocols. We find power law scaling $\tau_{th} \sim (1/T)^{(2n+2)}$, where $n$ is the order of the RMD. (b) The thermalization time scaling exponent $\alpha$ as a function of the initial state energy density. The results are similar to the $h\ne 0$ case.}
    \label{fig:no_long_field}
\end{figure*}

Our analysis so far has focused on the situation, where the longitudinal field $h$ is non-zero. In fig.~\ref{fig:no_long_field}, we show that the thermalization time scaling also holds when $h=0$. In this context, it is worth noting that for spin$-1/2$ systems, the thermalization time scaling exponent, $\alpha$ depends sensitively on the presence or absence of $h$; in particular, $\alpha = 2n+1$ when $h\ne 0$ and $\alpha = 2n-3$, when $h=0$~\cite{zhao2023temporal}. This result constitutes yet another difference between the dynamics of our classical spin model and its quantum counterpart.

\newpage

\section{Finite Size Effects}

\begin{figure*}[ht!]
    \centering
    \includegraphics[width=0.85\textwidth]{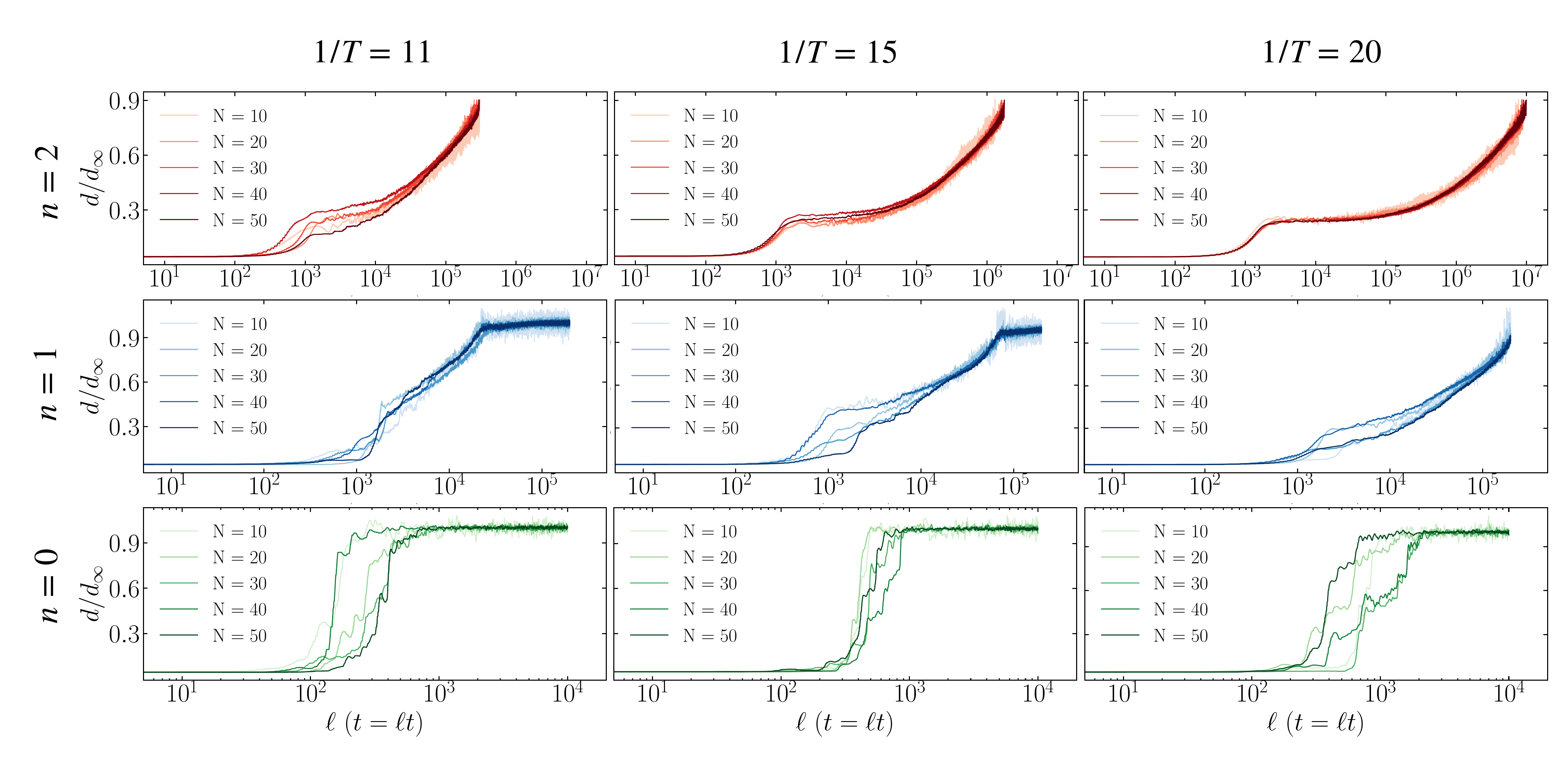}
    \caption{The evolution of the decorrelator for five different system sizes ($N \cross N$) at frequencies $1/T = 11, 15,$ and $20$. We observe that at higher frequencies for higher-order drives, the thermalization times remain roughly independent of the system size.}
     \label{fig:finite_effects}
\end{figure*}

We check for finite-size effects by looking at the decorrelator for various system sizes at a few frequencies for $n= 0, 1, 2$ RMD protocols. As shown in fig.~\ref{fig:finite_effects}, we find that for higher-order RMD protocols the thermalization time remains roughly system-size independent.  

\section{Stability of the Time Rondeau Crystal}

\begin{figure*}[ht!]
    \centering
    \includegraphics[width=0.75
\textwidth]{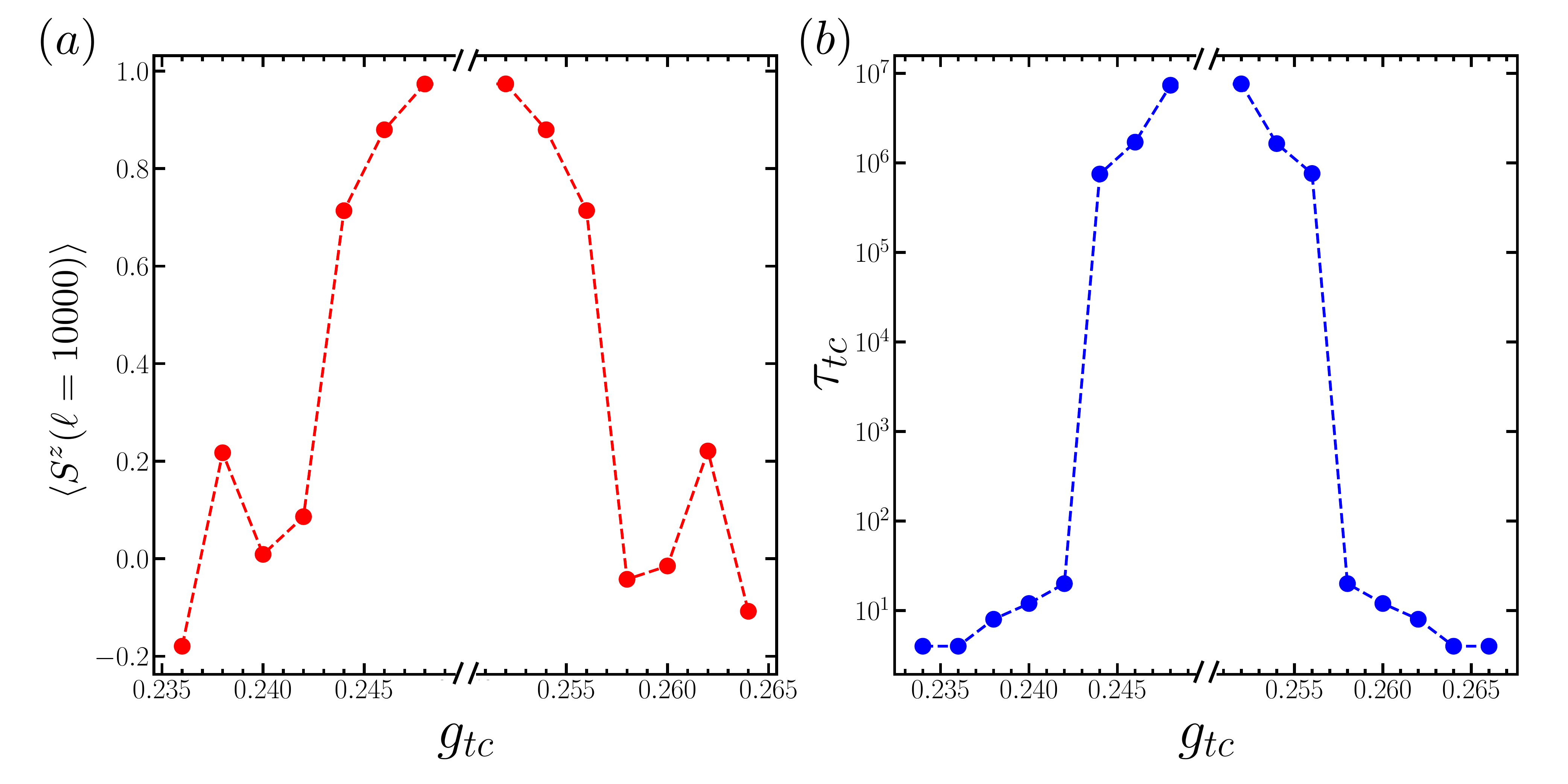}
    \caption{Phase diagram of the Time Rondeau Crystal: (a) The long-time value of the magnetization: $\langle S^{z} \rangle$ computed at $t= 10^4 T$, and (b) The lifetime, $\tau_{tc}$ of the time rondeau crystal as a function of $g_{tc}$. We find that time crystalline order is present for a long time when $g_{tc} \in [0.244, 0.256]$.}
    \label{fig:time_crystal_stability}
\end{figure*}

In the main text, we have demonstrated that our model exhibits long-time periodic temporal order in the presence of random multipolar driving when $g_{tc} = 0.255$; this kind of novel non-equilibrium order is dubbed a `time rondeau crystal'~\cite{moon2024experimental}. In this section, we check the stability of this phase around $g_{tc} = 0.25$ for the Thue-Morse drive. We do this analysis in two complementary ways. First, we examine the value of the order parameter (in this case, the magnetization, $\langle S^{z} \rangle$) at long times $(t=10^4 T)$ and find that $\langle S^{z} \rangle$ remains large as long as $g_{tc} \in [0.244, 0.256]$ (see fig.~\ref{fig:time_crystal_stability} (a)). Next, we compute the lifetime of the time crystal by determining the time, $\tau_{tc}$ at which the stroboscopic magnetization, $S^{z}(4lT)$ falls below a critical value, $S^{z}_{\rm cr}$. As shown in fig.~\ref{fig:time_crystal_stability}(b), a long-lived time-crystalline response can be seen when $g_{tc} \in [0.244, 0.256]$; we have set $S^{z}_{\rm cr} = 0.25$ for this calculation. Our analysis establishes the stability of time rondeau crystal phase in classical many-body systems.

\section{Robustness of the scaling exponents}

\begin{figure*}[ht!]
    \centering
    \includegraphics[width=0.5\textwidth]{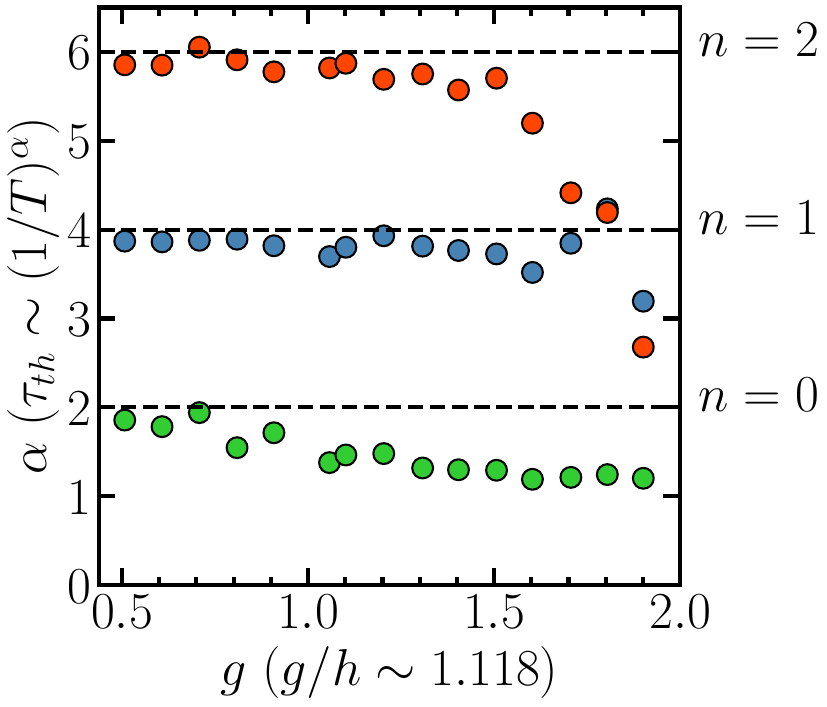}
    \caption{The thermalization time scaling exponent $\alpha$ for different values of $g$ and $h$ (with $g/h = 0.9045/0.809 \sim 1.118$). The $2n+2$ scaling for the exponent is followed for a wide range of parameter values (up to $g \sim 1.5$).}
    \label{fig:robust_exponents}
\end{figure*}

In the main text, we determined the thermalization time scaling exponents for $g=0.9045$ and $h=0.809$; this choice was made to connect to well-known results on Floquet thermalization~\cite{howell2019asymptotic}. In this section, we provide evidence of the robustness of these thermalization time exponents. We do this by varying $g$ over a large range, keeping the ratio $g/h = 0.9045/0.809 \sim 1.118$ fixed. As shown in fig.~\ref{fig:robust_exponents}, the thermalization time, $\tau_{\rm th}$ follows the same power law scaling $\tau_{th} \sim (1/T)^{\alpha}$ with $\alpha \sim 2n+2$ for $n$-RMD for a wide parameter regime (up to $g \sim 1.5$), thereby demonstrating the robustness of the results provided in the main text. 

\section{An equivalent quantum spin model}

\begin{figure*}[]
    \centering
    \includegraphics[width=0.9\textwidth]{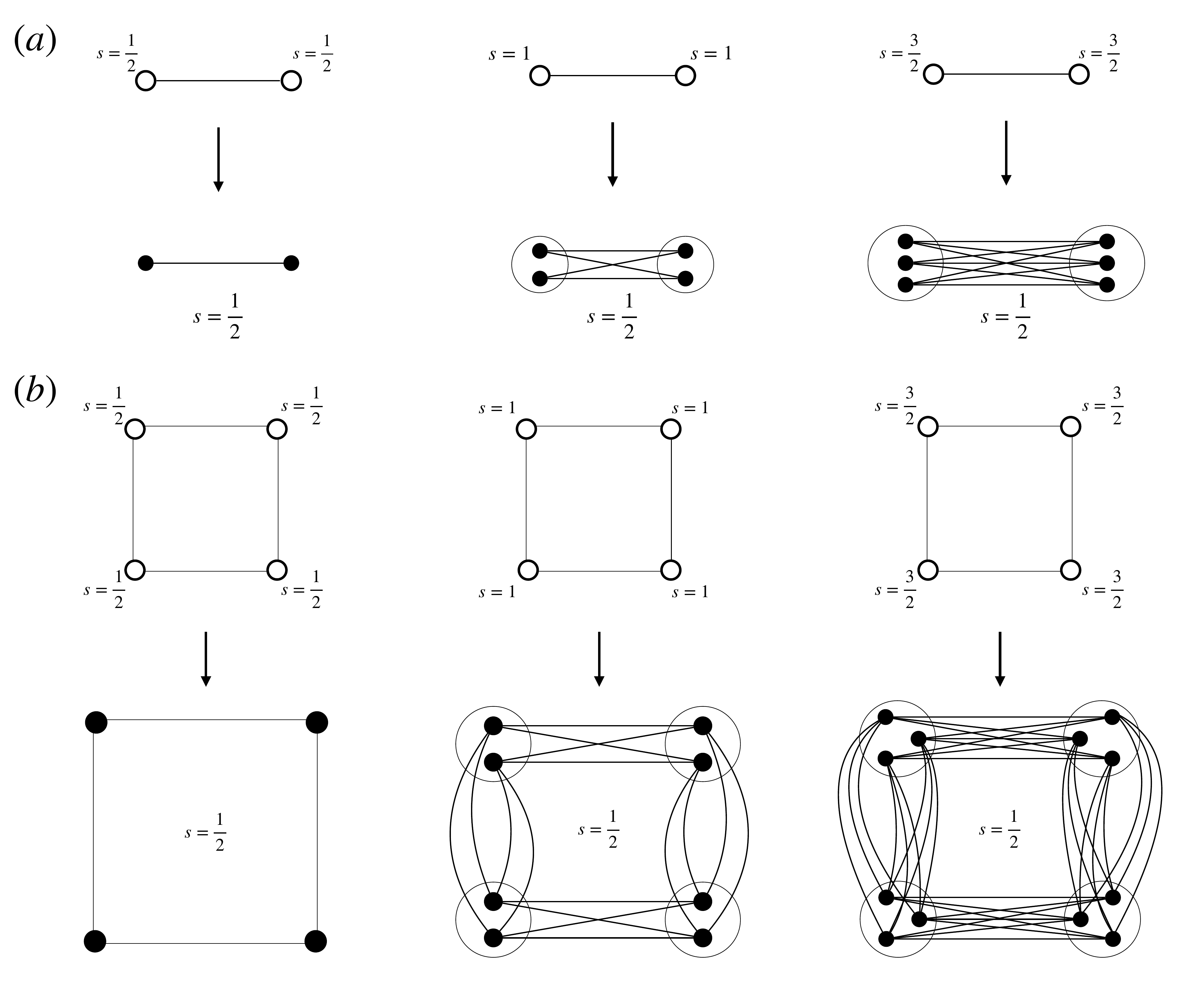}
    \caption{An illustration for mapping a spin-$S$ model to a quantum spin-$1/2$ model with an extra dimension. The effective spin-$1/2$ model has long-range interactions along this extra dimension.}
    \label{fig:quantum_model_illustration}
\end{figure*}

In the main text, we have stated that the dynamics of the classical spin model is equivalent to a long-range interacting quantum model composed of interacting spin-$1/2$ particles. We now proceed to demonstrate this equivalence. This section largely follows the prescription in ref.~\cite{mori2018floquet}. We begin by noting that the classical model that we have discussed arises in the $S \rightarrow \infty$ limit of a spin-$S$ quantum model. For our discussion, the relevant spin-$S$ Hamiltonians corresponding to $H_z$ and $H_x$ are
\begin{eqnarray}
    \label{hz}
    H_z &=&  \frac{1}{4S}\sum_{\bf r, r^{\prime}}  S^z_{\bf r} S^z_{\bf r^{\prime}} + \sum_{\bf r} h S^z_{\bf r}, \\
    \label{hx}
    H_x &=& \sum^{\bf r} gS^x_{\bf r},
\end{eqnarray},
where $|{\bf r-r^{\prime}}| = 1$.\\

Let us now consider the time-evolution of the system from an initially unentangled state:
\begin{equation}
 \vert \Psi(t=0)\rangle=\bigotimes_{\bf r} |\psi_{\bf r}(t=0)\rangle
\end{equation}
Furthermore, let us consider a classical initial state such that $\langle \Psi(t=0) \vert  S_{\bf r} \vert \Psi(t=0) \rangle^2 = S^2 $, where $S^2$ is the maximum possible value. In the $S \rightarrow \infty$ limit, the normalized spin vectors, ${\mathbf s_{\bf r}} = (\langle \Psi(t) \vert {\mathbf S_{\bf r}} \vert \Psi(t) \rangle)/S $, then evolve according to the equation:
\begin{equation}
    \frac{d}{dt} s_{{\bf r}}(t)  = \{s_{\bf r}, H\},
\end{equation} 
where $\{\ldots\}$ indicate Poisson brackets, and $\{s_{{\bf r}}^{\alpha},\, s_{{\bf r^{\prime}}}^{\beta}\} = \delta_{{\bf r} {\bf r^{\prime}} }\varepsilon^{\alpha\beta\gamma}s_{\bf r}^{\gamma}$. For a detailed proof of this statement, we refer the reader to Appendix A of ref.~\cite{mori2018floquet}.\\

We now note that we can decompose the spin-$S$ operator $S_{\bf r}$ into $2S$ spin-$1/2$ operators as follows: 
\begin{equation}
  S^{\alpha}_{\bf r} = \frac{1}{2}  \sum_{a=1}^{2 S} \sigma^{\alpha}_{{\bf r},a},
\end{equation} 
where ${\bf S}_{\bf r}^2 = S(S+1)$ and $\sigma^{\alpha}_{{\bf r},a}$ represent the standard Pauli matrices. The Hamiltonians $H_z$ and $H_x$ can then be written as:

\begin{eqnarray}
    \label{hzn}
    H_z &=&  \frac{1}{4S}\sum_{({\bf r}, a), ({\bf r^{\prime}},b)}  \sigma^z_{{\bf r},a} \sigma^z_{{\bf r^{\prime}},b} + \frac{h}{2} \sum_{({\bf r}, a)} \sigma^z_{{\bf r}, a}, \\
    \label{hxn}
    H_x &=& \frac{g}{2}\sum^{({\bf r}, a)} \sigma^x_{{\bf r}, a},
\end{eqnarray}

Thus, each site ${\bf r}$ in the spin-$S$ model is decomposed into site $({\bf r},a)$ in the spin-$1/2$ model; here $a$ ranges from $1$ to $2S$. The interactions amongst the spin-$S$ particles lead to an equivalent long-range interaction amongst the spin-$1/2$ particles along this extra dimension. An illustration of this mapping is shown in fig.~\ref{fig:quantum_model_illustration}. \\

We conjecture that this effective long-range interaction lead to heating suppression in the large-$S$ limit, thereby leading to longer prethermalization times. We attempted to validate this conjecture by studying the thermalization time for the effective long-range quantum model for $S= 1/2, 1,$ and $3/2$ on a $2\cross2$ square lattice. For these small-$S$ systems, we are unable to quantitatively determine the thermalization time scaling exponents for a $n-$RMD protocol. Our results on classical prethermalization indicate that for sufficiently large values of $S$, the thermalization time for a $n-$RMD would scales as $1/T^{(2n+2)}$. Unfortunately, this large-$S$ regime is beyond the reach of the exact diagonalization computations that we perform.
\bibliographystyle{apsrev4-2}
\bibliography{ref}